\documentclass[aip,amsmath,amssymb,reprint]{revtex4-1}
\usepackage[english]{babel}
\usepackage{physics}
\usepackage{graphicx}% Include figure files
\usepackage{dcolumn}% Align table columns on decimal point
\usepackage{bm}% bold math
\usepackage{stfloats}
\usepackage[normalem]{ulem}
\usepackage[draft,nomargin,inline,author=]{fixme}
\fxsetup{theme=color}
\definecolor{fxwarning}{rgb}{0.8,0.0000,0.0000}

\usepackage[utf8]{inputenc}
\usepackage[T1]{fontenc}
\usepackage{mathptmx}

\begin{document}

\preprint{AIP/123-QED}

\title[Emerging Materials for Neuromorphic Computing]{ITO-based Electro-absorption Modulator for Photonic Neural Activation Function}

\author{R. Amin}\author{J. George} \author{S. Sun} 
 \affiliation{Department of Electrical and Computer Engineering, George Washington University, Washington DC, 20052, USA.}
\author{T.~Ferreira~de~Lima} \author{A.~N.~Tait} 
 \affiliation{Department of Electrical Engineering, Princeton University, Princeton, NJ, 08544, USA}
\author{J. Khurgin}
 \affiliation{Department of Electrical and Computer Engineering, Johns Hopkins University, Baltimore, Maryland 21218, USA}
\author{M. Miscuglio}
 \affiliation{Department of Electrical and Computer Engineering, George Washington University, Washington DC,
20052, USA.}
\author{B.~J.~Shastri}
 \affiliation{Department of Electrical Engineering, Princeton University, Princeton, NJ, 08544, USA}
 \affiliation{Department of Physics, Engineering Physics \& Astronomy, Queen's University, Kingston, ON KL7 3N6, Canada}
\author{P.~R.~Prucnal}
 \affiliation{Department of Electrical Engineering, Princeton University, Princeton, NJ, 08544, USA}
\author{T. El-Ghazawi} 
\author{V. J. Sorger}
 \email{sorger@gwu.edu}
 \affiliation{Department of Electrical and Computer Engineering, George Washington University, Washington DC,
20052, USA.}
 
\date{\today}
\begin{abstract}
Recently integrated optics has become an intriguing platform for implementing machine learning algorithms and in particular neural networks. Integrated photonic circuits can straightforwardly perform vector-matrix multiplications with  high efficiency and low power consumption by using weighting mechanism through linear optics. Although, this can not be said for the activation function which requires either nonlinear optics or an electro-optic module with an appropriate dynamic range. Even though all-optical nonlinear optics is potentially faster, its current integration is challenging and is rather inefficient.
Here we demonstrate an electro-absorption modulator based on an Indium Tin Oxide layer, whose dynamic range is used as nonlinear activation function of a photonic neuron. The nonlinear activation mechanism is based on a photodiode, which integrates the weighed products, and whose photovoltage drives the elecro-absorption modulator.
The synapse and neuron circuit is then constructed to execute a 200-node MNIST classification neural network used for benchmarking the nonlinear activation function and compared with an equivalent electronic module. 
\end{abstract}

\maketitle
%\begin{quotation}
%\end{quotation}

\section{\label{sec:level1}Introduction}
With the ongoing advancement of neural network  systems, there is a pressing demand of new technological paradigms that can perform advanced artificial intelligence tasks without trading off throughput (operations per second) and power dissipation. Photonics based artificial neurons can potentially provide the archetypal solution for this specific challenge.\cite{prucnal_neuromorphic_2017,shastri_principles_2018,shen_deep_2017} The main advantage of photonics over current digital electronics implementations is that distinct signals due to their wave-nature can be straightforwardly and efficiently combined exploiting attojoule efficient electro-optic modulators\cite{heni_plasmonic_2019,amin_attojoule-efficient_2018,nozaki_ultralow-energy_2017,chiu_design_2015}, phase shifters and combiners, simplifying essential operations such as weighted sum or addition, vector matrix multiplications or convolutions.\cite{ferreira_progress_2017} 
Moreover, photonics enables high parallelism\cite{brunner_parallel_2013}, hence a higher baud-rate, since multiple wavelengths can travel in the same physical channel by exploiting wavelength-division multiplexing (WDM)\cite{blumenthal_integrated_2018,moscoso-martir_calibrated_2017,mehrabian_pcnna:_2018,tait_broadcast_2014,ferreira_progress_2017}. Additionally, photonics is marked by a further degree of freedom in modulating the information carried by the optical waves, since a signal can be modulated by altering its phase, amplitude or polarization. 

In a photonic neural network, a photonic neuron performs the following operations: receive multiple input signals (fan-in); weight each by coefficient and sum them (weighted sum or addition); followed by nonlinear process (nonlinear activation function); and sends its output to other neurons or nodes (fan-out). The nonlinear activation function has a key role in the network pivotal to the convergence into final states, by discriminating data and suppressing noise. Different nonlinear activation functions with significantly different ranges and trends \cite{george_neuromorphic_2019} have been proposed and extensively investigated, providing suitable advantages according to the different applications. More in details, interesting experimental and numerical all-optical nonlinear module based on saturable and reverse absorption \cite{miscuglio_all-optical_2018,dejonckheere_all-optical_2014}, graphene excitable lasers \cite{ma_all-optical_2017,shastri_spike_2016}, two-section distributed-feedback (DFB) lasers\cite{peng_neuromorphic_2018}, quantum dots\cite{mesaritakis_artificial_2016}, disks lasers\cite{alexander_excitability_2013, coarer_all-optical_2018}, induced transparency in quantum assembly \cite{miscuglio_all-optical_2018} have recently been reported and showed promising results in terms of efficiency and throughput for different kinds of neural network and applications, ranging from convolutional neural network, spiking neural network and reservoir computing. Although, a more straightforward implementation is currently attained by exploiting electro-optic tuned nonlinear materials \cite{cai_electrically_2011,kauranen_nonlinear_2012} or absorptive modulator directly connected to a photodiode, as shown in\cite{tait_neuromorphic_2017,george_neuromorphic_2019,george_electrooptic_2018,tait_neuromorphic_2017-1,yoo_electro-optical_2016}. In this case, the photogenerated current, proportional to the detected optical power at the weighted addition, alters the voltage drop on the active material, thus changing its carrier concentration and consequently the effective modal index of the propagating waveguide mode. This approach is affected by RC-latency and by the electro-optic conversion, therefore it trades off baud-rate with energy efficiency \cite{amin_waveguide-based_2018}, nonetheless it still provides higher controllability, reconfigurability, reliability and easy-integration with respect to more exotic design choices.  In this case, the adoption of a specific modulator type will directly impact the shape of the nonlinear activation function, subsequently the inference mechanism and ultimately the performance of the neural network. 
Indeed, the performance of a modulator is directly related to the underlying physics of the entirety of the opto-electronic devices, including active material and device configuration. Therefore, for achieving efficient modulation, and consequently, promising performance for the neural network, one should concurrently engineer both. 
On the device engineering side, maximizing the fraction of the optical field in the mode overlapping with the active material, namely intensifying the modal confinement factor, is a suitable and accessible strategy for enhancing the modulation efficiency and dynamic range of an electro-optic modulator embedded in a photonic integrated circuit. For achieving that, current schemes consist of using sub-diffraction limited plasmonic structures \cite{oulton_hybrid_2008,sorger_ultra-compact_2012,haffner_low-loss_2018,keeler_multi-gigabit_2017,zhu_phase_2013,wu_plasmonic_2008,gan_high-contrast_2013,briggs_efficient_2010} or photonic cavities \cite{gan_high-contrast_2013} aiming to maximize the light matter interaction in order to achieve a rather high modulation performance and low energy-per-compute surpassing electronic efficiency, while compensating in terms of insertion losses (optical mode hybridization) due to the plasmonic nature of the mode. Another characteristic to deem when engineering an electro-absorption modulator is the modulation bandwidth, which is the result of material choices and device configuration, and if well engineered, can enable high-throughput communication links. 
Though necessary, a thorough device configuration might not be enough for achieving an efficient complex index modulation due to inefficient active material choices. The main aspect to consider when designing and engineering an effective EA modulator is, in fact, the variation of the complex refractive index due to applied bias (i.e. carrier tunability), which is inherent to the selected active material\cite{george_electrooptic_2018}. Silicon (Si) is the conventional material choice usually as fabrication facilities can benefit tremendously from the mature Si process. But the inherent low tunability of Si under electrical bias forces inadequate performances at higher scaling as increased  modulator lengths need to be employed to achieve desired dynamic range for the nonlinear activation function. 

Moreover, in order to respond to the demand of densely integrated neural networks, the material of choice for the EA modulator needs to be easily and consistently integrated in a photonic platform; and preferably, the selected active material should be CMOS compatible, thus allowing integration with on-chip memory and EO converter such as digital-to-analog converter (DAC) and analog-to-digital converter (ADC) enabling large-scale functional photonic network-on-chip.
Undoubtedly, one class of materials which adapts to these requirements is transparent conductive-oxides (TCOs). Indium Tin Oxide (ITO) belongs to this class of material. The advantage of using ITO as active material in EA modulators are manifolds. ITO films are able to deliver unity-strong index modulation, when placed in a plasmonic cavity configuration \cite{sorger_ultra-compact_2012,lee_nanoscale_2014,vasudev_electro-optical_2013,babicheva_transparent_2015} and also show also strong modulation in epsilon-near-zero (ENZ) behavior\cite{abdelatty_hybrid_2018} in the telecommunication frequency band \cite{alam_large_2016}, supporting both strong index modulation and slow light effects\cite{zhao_silicon_2014}. In the recent years, the massive demand of ITO for industrial purposes \cite{noauthor_us8049862b2_nodate} and research-level advancements favored the rapid development of controllable and tunable processes for ITO deposition, by RF \cite{kerkache_physical_2006,chityuttakan_influence_nodate,hu_effects_2004,bhatti_characterization_2004,nozaki_ultralow-energy_2017} and DC sputtering\cite{lee_behaviors_2004}, and evaporation \cite{amin_0.52_2018}, which enabled a very high yield and reliable process and monolithic hybrid integration \cite{gui_towards_2018}. 
In this paper, we start our study by analyzing the free carrier absorption dynamics in ITO (Sec. \ref{sec:theory}), only to then build a model of a capacitively coupled electro-optic neuron for the elctro-absorption modulator (Sec. \ref{sec:sim}). Subsequently, we derive its cascaded signal-to-noise ratio (SNR) as function of its length and laser power. We compare the model obtained with Silicon-based modulator, being still largely employed with successful results \cite{tang_over_2012,tu_fabrication_2011,soref_electrooptical_1987,reed_recent_2013,zhu_phase_2013,haffner_all-plasmonic_2015}. (Sec. \ref{sec:sim}) 
As a sub-case of our study, we also experimentally demonstrate an ITO-based absorption modulator, which due to intrinsic nonideality such as initial doping and asymmetric behaviour outperforms the theoretical model initially described. Ultimately, we evaluate and benchmark the performance of the experimentally fabricated ITO modulator as nonlinear activation function on the well-established MNIST feed-forward neural network, designed with two layers of 100 nodes each, using optimized parameters from the SNR analysis. The value of accuracy for the trained network, including noise, reached a remarkable 98\%, exploiting also non-ideal and asymmetric behavior in the transfer function, discussed in the results section (Sec. \ref{sec:device}). 
In conclusion, our results show that ITO-based electro-absorption modulators offer significantly higher modulation dynamic range and efficiency with respect to similar Silicon-based devices; and hence, are quite suitable for implementing the nonlinear activation functions in electro-optic neural networks as they require only a small fraction of the power and footprint. We believe that this work could constitute a viable approach not only towards the implementation of a photonic perceptron mechanism based on ITO modulators, both for weighting the inputs and implementing nonlinear activation functions, but also could represent a comprehensive guide for other material integration to avail future neural network trends.

\section{Methods \label{sec:theory}}
Initially, we study the electro-optical absorption dynamics in ITO dictated by the free carrier mechanism, which is essential for determining the model for the absorption as function of the injected carriers and bias voltage. We compare these results with Silicon carrier dynamics as the underlying modulation mechanism for both Si and ITO are the same - free carrier absorption arising from accumulation/depletion of the carriers.
\begin{figure}[h]
\includegraphics[width=0.44\textwidth]{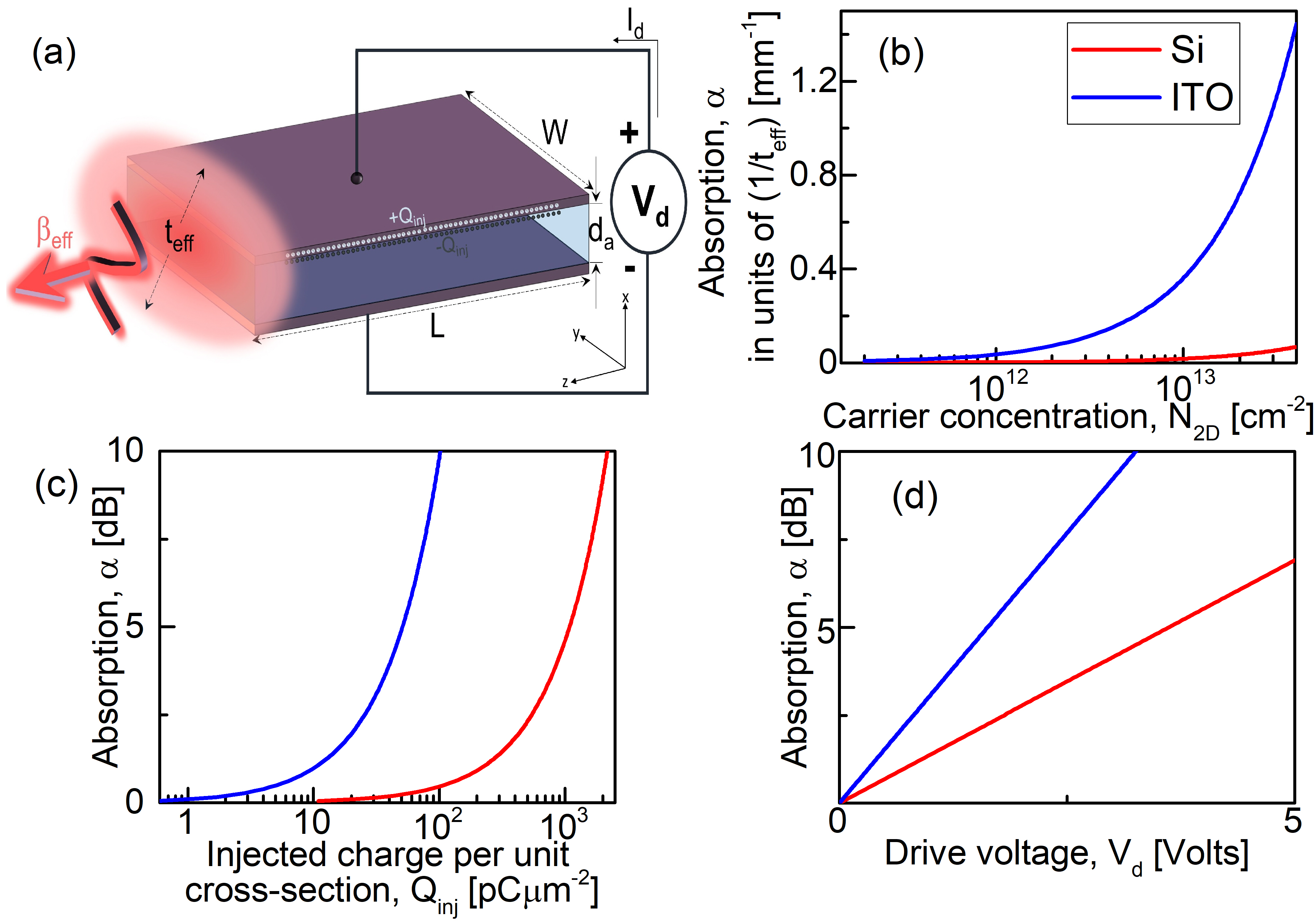}% Here is how to import EPS art
\caption{Free carrier based absorption modulators in Silicon (Si) and Indium tin oxide (ITO). (a) Schematic of a generic waveguide modulator showcasing the cross-sectional structure (width, $W$ and active layer thickness, $d_a$) and modal effective thickness, $t_{\text{eff}}$. The mode inside the structure is propagating in the z-direction with propagation constant, $\beta_{\text{eff}}$; $L$ is the length of the modulator; $V_d$ and $I_d$ are the drive voltage and drive current, respectively, responsible for the injected charge accumulation/depletion ($\pm Q_{\text{inj}}$) in the active layer/gate interface. (b) Absorption, $\alpha$ in units of $1/t_{\text{eff}}$ [mm$^{\text{-1}}$] of our cross-sectional geometry in (a) vs. 2-dimensional planar carrier concentration, $N_{\text{2D}}$ [cm$^{\text{-2}}$]. (c) Absorption, $\alpha$ [dB] vs. charge injected per unit cross-sectional area, $Q_{\text{inj}}$ [pC$\mu$m$^{\text{-2}}$]. A 10 dB ceiling for the attainable extinction ratio was chosen arbitrarily. (d) Absorption, $\alpha$ [dB] vs. drive voltage, $V_d$ [Volts] corresponding to electrostatic gating in the presence of a gate dielectric with $\epsilon_{\text{eff}}$ =10 and thickness, $d_{\text{spacer}}$ = 5 nm.}\label{fig:epsart}
\end{figure}

\subsection{Free carrier absorption dynamics}\label{sec:theory2}
Let us consider a free electron gas in the conduction band of a semiconductor such as in Si, ITO, or any other similar material with the carrier density of  $N_c (x,y)$. The schematics of a generic waveguide modulator case is shown in Fig. \ref{fig:epsart}(a) employing such active materials. Here, we derive the expression for the absorption cross-section of free electrons in these akin materials. The relative permittivity of aforementioned free electron plasma in the Drude approximation can be written as 
\begin{equation}
    \varepsilon_r(\omega,x,y)=\varepsilon_{\infty}-\frac{N_c(x,y)q^2}{\varepsilon_0m_{\text{eff}}}\frac{1}{\omega^2+i\omega\gamma}
\end{equation}
where $\varepsilon_{\infty}$ is the dielectric constant of the undoped semiconductor also referred to as the high-frequency or 'background' permittivity, $q$ is the electronic charge, $m_{\text{eff}}$ is the conduction effective mass, $\omega$ is the angular frequency, and $\gamma$ is the collision frequency of the free carriers quantified by the detuning from any interband resonances thereof. The rate of absorption of electro-magnetic energy is proportional to the imaginary part of the dielectric constant,   as
\begin{equation}
\begin{split}
\frac{\text{d}U(x,y)}{\text{d}t}=&\omega\varepsilon_0\varepsilon_{r,\text{im}}(\omega,x)\frac{E^2_{\text{eff}}(x)}{2}\\=&\frac{N_c(x,y)q^2}{m_{\text{eff}}}\frac{\gamma}{\omega^2+\gamma^2}\frac{E^2(x,y)}{2}
\end{split}
\end{equation}
where the effective electric field, $E_{\text{eff}}$  is the same as the total field, $E$. Now, let us consider an electromagnetic wave in a waveguide with width $W$, propagating along the z-direction (Fig. \ref{fig:epsart}(a)), with the propagation constant $\beta$ expressed as $E=E(x,y)e^{i(\beta z-\omega t)}$. We can evaluate the Poynting vector as $S_z(x)=E_y H_x$. According to Maxwell equations $\frac{\partial{H_x}}{\partial z}=\beta H_x=\omega n^2 (x)\epsilon_0E_y$. Thus, the time-averaged Poynting vector magnitude is $S_z(x)=\omega n^2(x)\varepsilon_0E^2_y/2\beta=n^2(x)E_y^2/2n_{\text{eff}}\eta_0$, where $\eta_0$ is the impedance of free space ($\simeq $377$\Omega$). The effective index has been introduced as $n_{\text{eff}}=\beta c/\omega$. Integrating over the waveguide cross-section (Fig. \ref{fig:epsart}(a)) we obtain

\begin{equation}
\begin{split}
\dv{P}{z}=&-\int_{\text{active}}\int_{-W/2}^{W/2}\dv{U}{t} \,dx\,dy\\  &= -\frac{q^2}{m_{\text{eff}}}
\frac{\gamma}{\omega^2 + \gamma^2} \frac{W}{2}\int_{-\infty}^{\infty}N_c(x)E^2(x)\,dx 
\end{split}
\end{equation}                          
The total power flow then becomes 
\begin{equation}
\begin{split}
    P=& W\int_{-\infty}^{\infty}S_z(x)\,dx=\frac{W}{2n_{\text{eff}}\eta_0}\int_{-\infty}^{\infty}n^2(x)E^2_y\,dx\\
    =& \frac{Wt_{\text{eff}}n_{eff}}{2\eta_0}\frac{\int_{-\infty}^{\infty}N_c(x)E^2(x)\,dx}{N_{2D}}
\end{split}
\end{equation}
Here, $N_{2D}$ is defined as the 2-dimensional planar carrier concentration and the effective thickness is 
\begin{equation}
\begin{split}
    t_{eff}=& \frac{1}{n^2_{eff}} 
    \frac {\int_{-\infty}^{\infty}N_c(x)\,dx\int_{-\infty}^{\infty}n^2(x)E^2_y\,dx}{\int_{-\infty}^{\infty}N_c(x)E^2(x)\,dx}\\
    \approx& \frac{1}{n^2_{eff}}\int_{-\infty}^{\infty}n^2(x)E^2_x\,dx/E^2_{a0}
\end{split}    
\end{equation}
For a very narrow active layer, this approximation can be inferred – otherwise it is slightly smaller. This definition of the effective thickness may differ from many other in the literature, but the difference is not significant. Usually it is just in the arrangement of all kinds of indices. Note, this definition includes the fact that active layer is not necessary at the center of the waveguide, i.e. $E_{a0}$ need not be the peak electric field. The effective thickness can also be thought of as a relative measure of the confinement of the optical mode to the active layer as confinement factor, $\Gamma \simeq d_a/t_{\text{eff}}$. Now, we can obtain from (3) as
\begin{equation}\label{eq:eq1}
\frac{dP(z)}{dz}=
-\frac{q^2\eta_0}{m_{eff}
n_{eff}}
\frac{N_{2D}}{t_{eff}}
\frac{\gamma}{\omega^2+\gamma^2} P(z)
= -\alpha(\omega)P(z)
\end{equation}
The absorption coefficient can be found as $\alpha(\omega)=\sigma_{abs}(\omega) N_{2D}/t_{eff}$ ; where the absorption cross-section is expressed as 
\begin{equation}\label{eq:eq2}
\begin{split}
\sigma_{abs}(\omega)=& 
\frac{q^2\eta_0}{m_{eff}n_{eff}\gamma}
L(\omega)
= \frac{4\pi\alpha_0}{n_{eff}}
\frac{\hbar}{m_{eff}}
\frac{\gamma}{\omega^2+\gamma^2}\\
=& \frac{4\pi\alpha_0}{n_{eff}}
\frac{\hbar^2}{m_0(\hbar\gamma_{eff})}
\frac{m_0}{m_{eff}}\\
\approx& \frac{7.02\times 10^{-17} cm^2}{\hbar\gamma_{eff}} \times \frac{m_0}{m_{eff}}
\end{split}
\end{equation}
Here, $\alpha_0$ is the fine structure constant, and $L(\omega)=\gamma^2/(\omega^2+\gamma^2)$ represents a Lorentzian line shape that has maximum value at resonance equal to unity. The effective detuning then becomes $\gamma_{eff}=\gamma +\omega^2/\gamma$ . Note, the maximum absorption is achieved at $\gamma=\omega$ and when $\gamma_{eff}=2\omega$ and
\begin{equation}\label{eq:eq3}
\begin{split}
\sigma_{abs}(\omega)|_{max} =& \frac{4\pi\alpha_0}{n_{eff}}
\frac{\hbar}{m_{eff}\omega} \times \frac{1}{2}\\
\approx& 3.53\times 10^{-17} cm^2
\frac{1}{\hbar \omega n_{eff}}
\frac{m_0}{m_{eff}}
\end{split}    
\end{equation}
Evidently, this expression is similar to the absorption cross section of semiconductor quantum dots (QDs) with one major difference – due to non-resonant character of the free carrier absorption, the cross section for the free carriers is orders of magnitude lower than that for the resonant QDs \cite{amin_waveguide-based_2018}. As one can see, absorption cross-section does not depend on too many things – at the resonance it depends only to resonance width, or, better $Q=\omega/\gamma$ and the optical dipole  transition rate, $r_{12}$. We can invoke the oscillator sum rule (assuming all the dipoles along the axis parallel to the electric field are lined up) which states $2m_{eff}\omega r^2_{12} \leq \hbar$. Therefore, we can introduce the oscillator strength as $f_{12}=2m_{eff}\omega r^2_{12}/\hbar \leq 1$ and consequentially get
\begin{equation}\label{eq:eq4}
\sigma_{abs} =2 \sigma_{max} \frac{\gamma^2}{\omega^2+\gamma^2} \approx 2 \sigma_{max}\frac{\gamma^2}{\omega^2}    
\end{equation}
If we now assume that the material used is a semiconductor with a bandgap, $E_g$ and the momentum matrix element quantified by $P_{cv}$, we can actually relate the effective mass of electron to the wavelength, 
\begin{equation}\label{eq:eq5}
\frac{m_0}{m_{eff}}=1+\frac{2P^2_{cv}}{m_o E_g} \approx \frac{E_p}{E_g}    
\end{equation}
where $E_p=2P^2_{cv}/m_0 \approx 16 - 24 eV$, for a wide range of semiconductors. Therefore, we obtain (assuming that the bandgap is close to the photon energy)  
\begin{equation}\label{eq:eq6}
\sigma(\omega) \approx 9.6\times 10^{-15} cm^2 \frac{f_{12}}{(\hbar \gamma)(\hbar \omega)n_{eff}}    
\end{equation}
Thus, while modulators based on two-level systems do inherently benefit from the small $\gamma$, the condition for the free carrier absorption based modulators is quite opposite as they benefit from for large $\gamma$. Also, we see that for the wavelength in the telecom range the cross-section actually scales with broadening caused by scattering $\gamma$. Therefore, ITO leads intrinsically to more modulation per unit charge than high-quality Silicon. We found in our previous work, that for all materials operating with Pauli blocking (saturable absorption schemes) show comparable values of absorption cross-section ($\sigma \sim 10^{-14} cm^2$ at room temperature)\cite{amin_waveguide-based_2018}, whilst the free carriers offer a worse performance due to non-resonant character of absorption, since according to (8) for free carriers $\gamma_{eff}=\gamma+\omega^2/\gamma > 2\omega$. To verify these analytical results, obtained from essentially perturbative approach, we calculate the dependence of the absorption coefficient on the injected (induced by the gate) carriers, $N_{2D}$   as the elemental free carriers responsible for absorption tuning in the electrostatic modulators as 
\begin{equation}\label{eq:eq7}
\alpha(\omega)= \frac{4\pi\alpha_0}{n_{eff}t_{eff}}
\frac{\hbar}{m_{eff}} \frac{\gamma}{\omega^2+\gamma^2}N_{2D}
\end{equation}
Here, $\gamma=1/\tau$ is the carrier scattering rate i.e. collision frequency with $\tau$ being the mean scattering time corresponding to the mean free path between collisions between the free electrons in the plasma. The electron mobility, $\mu$ and $\tau$ are related by $\mu=|q|\tau/m_{eff}$. We compare two free carrier absorption based materials for our absorption modulators to employ in the following neuromorphic study, namely Silicon and ITO. For Si, the conductivity effective mass, $m_{eff}$ is taken as $0.26m_0$, where $m_0$ is the free electron rest mass \cite{sze_physics_2006}. $\mu$ is taken as 1100 $cm^2V^{-1}s^{-1}$ at $10^{16} cm^{-3}$ carrier concentration level (i.e. electrons for Silicon)\cite{nashima_measurement_2001}. We used the dielectric constant of non-doped silicon $\epsilon_{\infty}$= 11.66\cite{holm_microwave_1968}. Unlike the doped silicon, the chemical composition of ITO is usually given as $In_2O_3:SnO_2$ and can be considered an alloy as concentration of Sn relative to In can be as high as 10\%. Several previous studies have calculated the permittivity of ITO using the experimentally measured reflectance and transmittance, and we chose a fitting result of Michelotti et. al whereas   depends on the deposition conditions, defect states, and film thicknesses, etc.; in our analysis we have taken $\gamma$ = 1.8$\times10^{14} rad s^{-1}$, background permittivity, $\epsilon_\infty$= 3.9, and $m_{eff} = 0.35m_0$ \cite{michelotti_thickness_2009,noginov_transparent_2011,vasudev_electro-optical_2013,naik_alternative_2013}. In the near-infrared, ITO is quasi-metallic since its free electrons dictate its optical response. In fact, ITO and related transparent conducting oxides have recently been explored as plasmonic materials in the optical frequency range\cite{noginov_transparent_2011,amin_low-loss_2018}.

The inherent low broadening parameter, $\gamma$ of Si limits the attainable absorption with carrier density level variation even at higher concentrations. More importantly, the slope of absorption vs. density characteristics is responsible for the modulator performance (i.e. slope steepness $ER/V_{bias}$), and as such ITO exhibits prominent features as the ITO curve is several orders per decade higher in steepness than the Si one (Fig. \ref{fig:epsart}(b)). Additionally, we evaluate the total absorption as a function of the injected charge per unit waveguide cross-sectional area setting a 10 dB ER ceiling, $Q_{inj} = qN_{2D}L/t_{eff}$ (Fig. \ref{fig:epsart}(c)). In accordance to our perturbative estimation the injected charge follows $Q_{inj} \sim 2.2q/\sigma$ \cite{amin_waveguide-based_2018}. Further, the drive voltage can be obtained via $V_d = Q_{inj}/C_g$, where $C_g$ is the gate capacitance (Fig. \ref{fig:epsart}(d)). A dielectric spacer, as such, an oxide layer for gating facilitation is considered in these analytical formalism. A gate oxide with a relative dielectric constant, $\epsilon_{eff}$ of 10 and gate spacer thickness, $d_{spacer}$ of 5 nm was chosen for acquiring these results. As the drive voltage is increased, free carriers leading to dispersive effects are induced and a corresponding net increase occurs in the carrier concentration.
While both ITO and Si are capable of tuning the index via the free carrier dependent Drude model, nevertheless ITO exhibits disparate significant advantages over Si with respect to the permittivity variation; firstly, the carrier concentration in ITO can exceed that of Si by at least couple of orders of magnitude; the concentration of indium atoms in ITO reaches a few percent, which is significantly beyond the attainable donor concentration in doped Si\cite{amin_deterministic_2017}. Secondly, the effect of inducing the carrier concentration change in ITO on its refractive index is more dramatic than in Si. This can be attributed to the higher bandgap and consequently lower refractive index of ITO compared to that of Si. The higher bandgap allows the density of states (DoS) near band edge to be tighter confined leading to a higher free carrier concentration necessary to change the index. For a discrete spectrum, the DoS consists of a number of delta peaks at the energy levels of the system with weighted degeneracy of that level. If the change of the carrier concentration $\partial N_{2D}$ (e.g. due to an applied bias) causes a change in the relative permittivity (dielectric constant) $\partial \epsilon$, the corresponding change in the refractive index can be written as $\partial n = \partial \epsilon ^{1/2} \sim \partial \epsilon /2 \epsilon ^{1/2}$, hence the refractive index change is greatly enhanced when the permittivity $\epsilon$ %%check MM, yes I think this should be in a proper math equation since the ^1/2 should only be in the demoninator.
is small\cite{amin_low-loss_2018,amin_0.52_2018}. Third, the presence of an epsilon-near-zero (ENZ) region in the tolerable carrier concentration range within electrostatic gating constraints makes ITO a formidable opponent to conventional Si. Operation in the vicinity of $\epsilon \sim 0$ (ENZ) condition can understandably result in stronger modulation effects as the optical mode experiences gradually intensifying slow-light effects closer to the ENZ region as a result of increasing light matter interaction thereof. The inherently low background permittivity of ITO allows the ENZ carrier density condition at telecommunication relevant wavelength of 1550 nm resulting in experimentally obtainable effects $(\simeq 6 - 7 \times 10^{20} cm^{-3})$ \cite{amin_low-loss_2018,amin_0.52_2018,amin_deterministic_2017}. Interestingly, Si ENZ condition can be approached at a carrier concentration level of about $3.38\times10^{19} cm^{-3}$, operating at a wavelength of 10 $\mu m$ in the IR region away from the region of our interest.

\subsection{\label{sec:sim}Neural Network}

\onecolumngrid
\begin{figure*}[!ht]
\includegraphics[width=.9\textwidth]{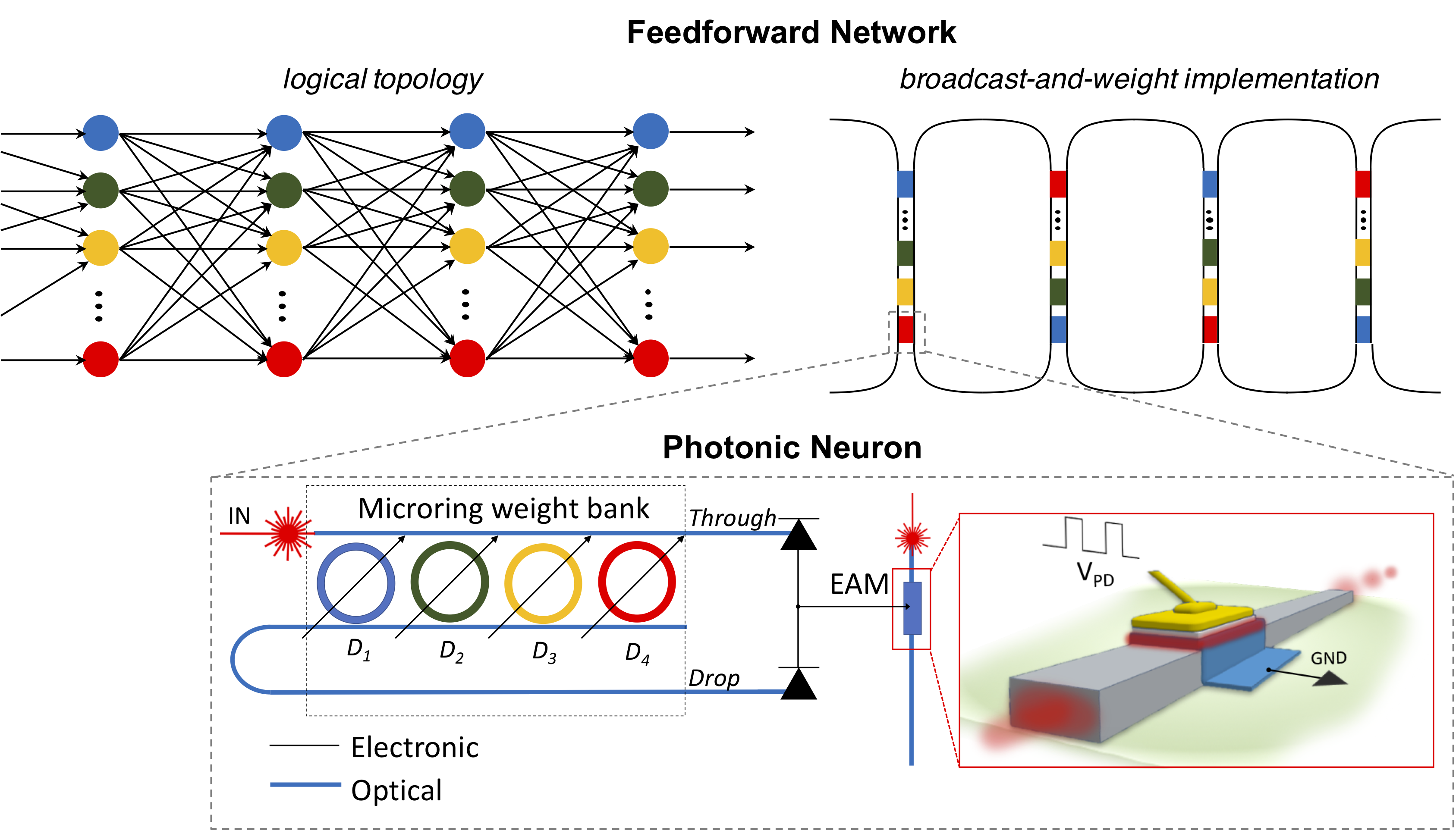}
\caption{\label{fig:fig2} Top: A photonic feedforward neurual network implementation using the broadcast-and-weight protocol. Here, feedforward networks can be constructed using interfacial photonic neurions that connect between broadcast mediums (waveguides) \cite{peng_neuromorphic_2018}. Bottom: Photonic Neuron and perceptron mechanism. WDM Inputs are weighed through tunable MRR. The optical power  is accumulated and detected by a balanced photodiode. The photo-voltage drives the electro-absorption modulator, which nonlinearly modulates the laser power mimicking an activation function}
\end{figure*}
\twocolumngrid

The model of the ITO absorption mechanism is then incorporated in the NL activation function of a  neural network. %\fxwarning{HTP: I think it's good to add a figure of input-output transfer function. This will make readers visualize the nonlinear activation function of a neuron.}
A standard broadcast-and-weight feed-forward network architecture in which connections are configured by microring weight banks \cite{peng_neuromorphic_2018}, characterized by two layers and 100 nodes per layer, is considered in Fig. \ref{fig:fig2}.
The network is trained to classify 10 individual digits in a set of images of handwritten digits in a grayscale 28$\times$28 pixel (MNIST dataset \cite{lecun-mnisthandwrittendigit-2010}). 
It is worth noticing, that here each pixel is encoded in unique wavelength carrier in a wavelength division multiplexed (WDM) and broadcast scheme.
Subsequently, the network inference performance are benchmarked. The simulation environment was developed in Python using Keras \cite{chollet2015keras} and TensorFlow \cite{tensorflow2015-whitepaper}. The activation function was replaced with a custom designed non linear transfer function resulting from our ab-initio ITO based electro-absorption modulator model. The weighting mechanism relies on microring resonator banks as proposed by Tait et al\cite{tait_neuromorphic_2017,tait_broadcast_2014}, which can potentially support more than 100 channels\cite{tait_broadcast_2014}. In this case, the weights were bound between minus one and one to simulate input optical weighting by rings in a push-pull configuration.

Noise is modeled in Keras as additive Guassian noise with standard deviation proportional to the RMS noise power. Shot noise at the photodiodes is modeled as $ \sqrt{2 q \left( I_s + I_d \right)\Delta f } $ Where $ I_s $ is the signal current and $ I_d $ is the dark current taken as 0.05 nA. In addition, thermal noise is modeled as $ \sqrt{4 k_B T \Delta f / R_{eq}} $, where $T$ is taken as room temperature of 300 K and $ R_{eq} $ is the equivalent resistance of $ 50\ \Omega $.

\subsubsection{Training}
The neural network models are trained in Keras \cite{chollet2015keras} with Adagrad \cite{Duchi:EECS-2010-24} method using a categorical cross-entropy loss function,  a learning rate of 0.005, zero decay, with 500 epochs of a 1024 batch size. The amplitude of the noise power model is reduced during training to 10\% of the final inference noise power, to allow for training convergence while keeping the model from overfitting to a noiseless activation function.

\subsubsection{Parameter Optimization}
The photonic neural network differs from the digital neural network in three ways: First, the noise of the photonic neural network is analog, with sources from shot and thermal noise, and unlike the digital neural network has no quantization noise. Second, the photonic neural network, as an analog system, cascades noise. However, unlike an analog repeater, the analog neural network acts as a signal regenerator when the activation function is pushed into saturation, making the analog neural network a quasi-digital system in certain configurations. Finally, the photonic neural network without parameter optimization will shift the bias and dynamic range moving with depth into the network.

To counter-act the shifting bias and dynamic range, an optimization procedure was designed to locate the center and range of the activation function at each layer of the network as follows: the activation function is swept over 200 input voltage points with no noise, the point with the greatest slope is found, a range is built by adding the points to the left and right of the starting point until the slope has dropped below 0.05. After identifying the center of the dynamic range, a DC bias is applied to the modulator to move the minimum input to the center of the identified range.
 
\begin{figure}[h]
\includegraphics[width=.45\textwidth]{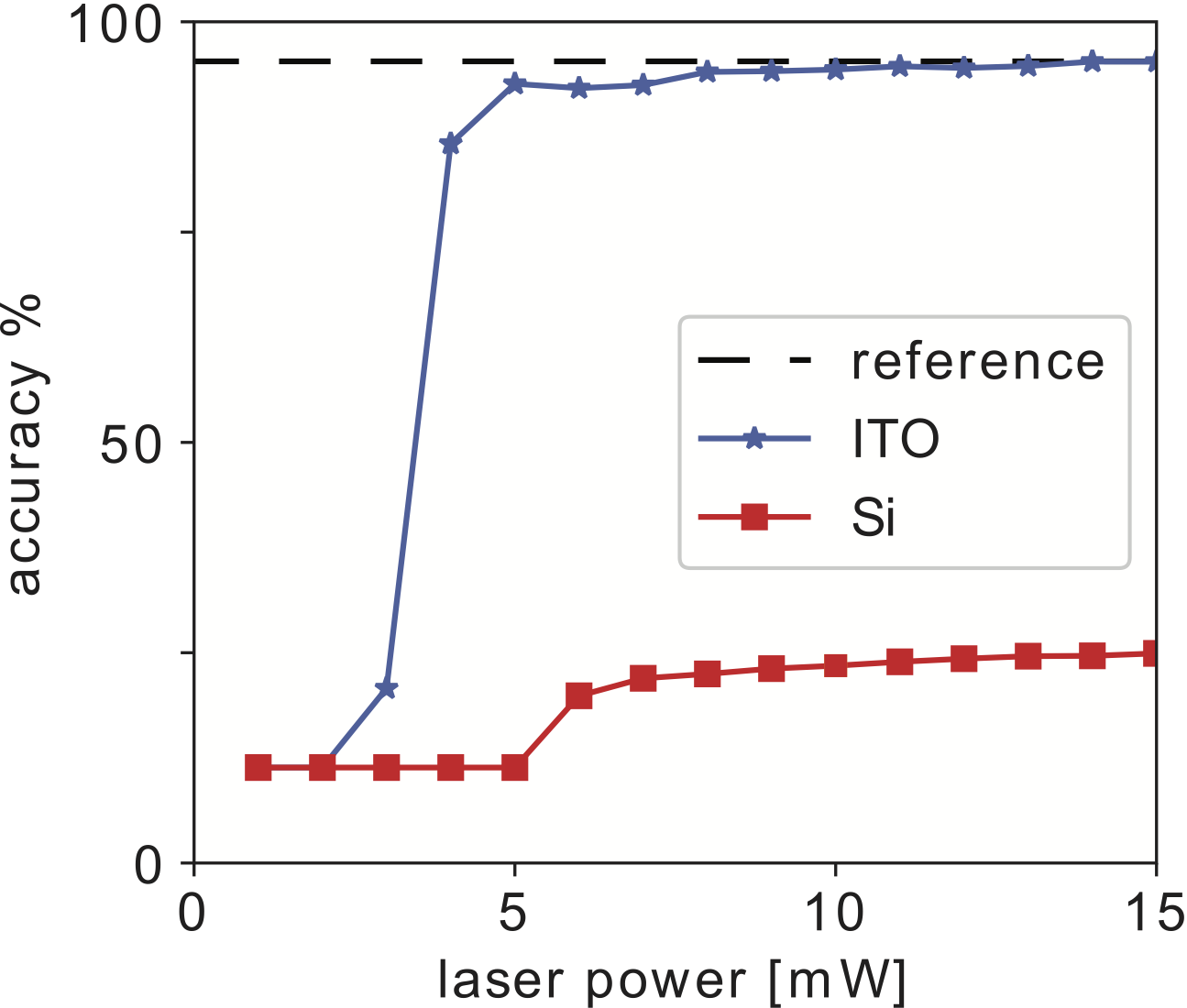}
\caption{\label{fig:ito_si_model_results}
Simulated MNIST accuracy results for ITO and Si modulator activation functions swept over laser power for a fully connected neural network of two layers each with 100 nodes.}%\fxwarning{Why is silicon so poor? --TFL}
\end{figure}

\section{\label{sec:results}Results and Discussion}
\subsection{\label{sec:sim_model}ITO electro-absorption activation function: Inference results}
An activation function derived from Eq. (\ref{eq:eq7}) for both ITO and Si based absorption modulators coupled to a photodiode (Fig. \ref{fig:fig2}) through an ideal transimpedance amplifier was trained in Keras\cite{chollet2015keras} at a simulated bandwidth of 1 GHz. The length for both modulators was taken to be the 10 dB dynamic range point for ITO at 1768 in units of the effective thickness, $t_{eff}$. We keep the scaling independent of the modal structure choice (only material dependent, i.e. Si/ITO) by using the effective thickness as the underlying unit for device length as $t_{eff}$ can be different for different modes but the physical length required for modulation relates back to $t_{eff}$ based on the modal choice as $L/t_{eff}$ depends only on material constraints (i.e. maximum attainable absorption)\cite{amin_waveguide-based_2018}. The simulation resulted in accuracy greater than 90\% for laser power greater than 5 mW (Fig. \ref{fig:ito_si_model_results}) for the ITO modulator, while the Si modulator never reached accuracy greater than 30\% even at the final sweep power of 15 mW. It is worth noticing that here we compare Si and ITO modulator with the same physical footprint exemplified by the modulator length in effective thickness ($\sim L/t_{eff}$) to achieve 10 dB modulation for the ITO device. Moreover, as aforementioned in section \ref{sec:theory2}, this effect is due to the lower modulation range ($ER/V_{bias}$) in Si compared to ITO  ascribable to the inherent low effective broadening parameter, $\gamma_{eff}$ of Si which drastically limits the attainable absorption even at higher carrier density levels. As such, similar accuracy level in  executing neural network tasks can be obtained for  Si based modulators with substantially larger linear footprint compared to ITO-based electro-absorption modulators, when used as nonlinear activation functions. Even with the enhanced footprint, Si based devices could be susceptible to accuracy challenges as opposed to their highly scalable ITO counterparts because of the steep dynamic response of the latter (Fig. \ref{fig:epsart}(d)).

\begin{figure}[h!]
\includegraphics[width=.48\textwidth]{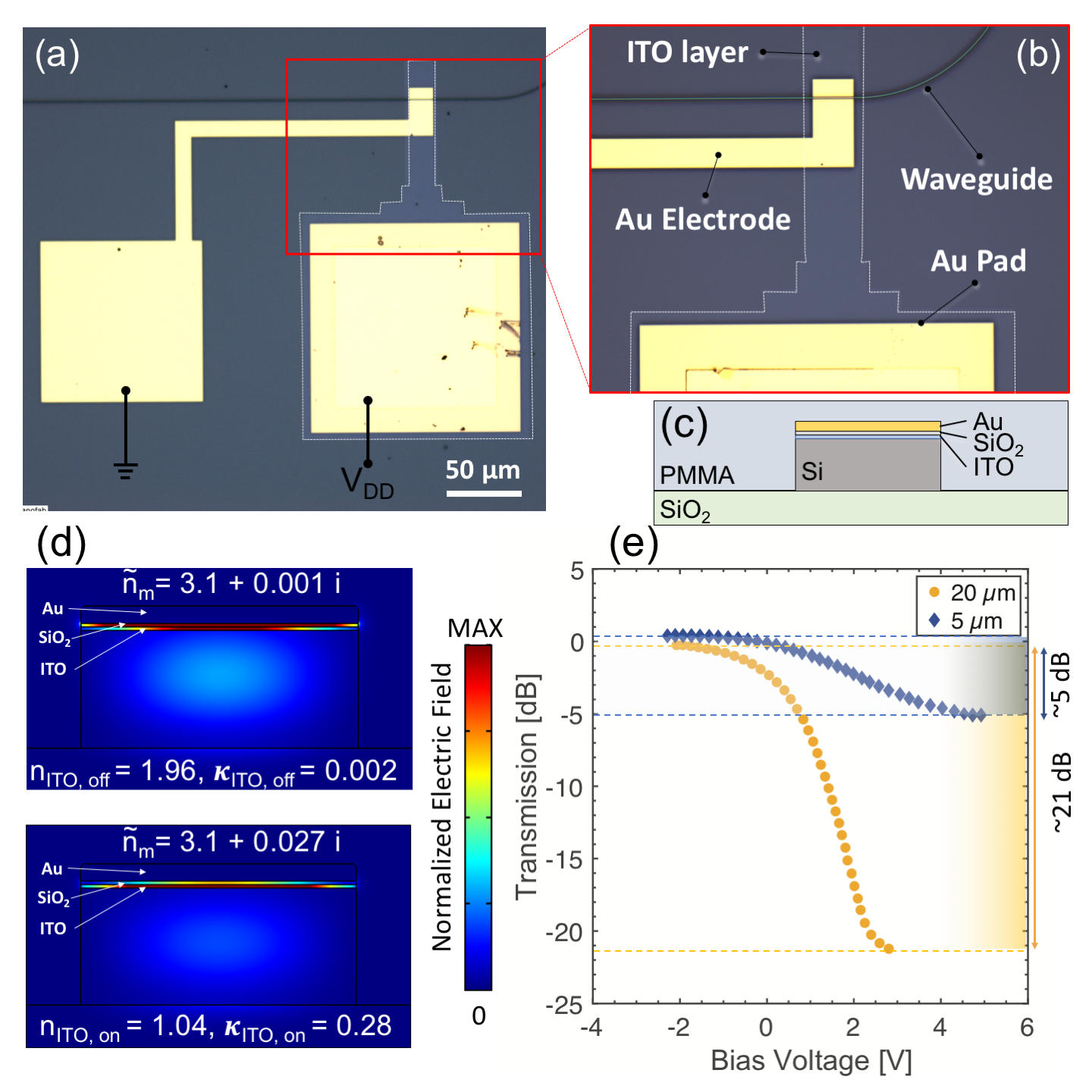}% Here is how to import EPS art
\caption{\label{fig:DEVICE} (a-b) Optical image of the electro-absorption device and schematic of the cross section (c). A 10 nm ITO film is deposited on a silicon waveguide (800 nm $\times$ 340 nm) covered by a thin SiO$_2$ spacer and a 40 nm Au metal contact. Scale bar is 50 $\mu$m. (d) Numerical simulation (FEM) of the normalized electric field distribution for the OFF and ON state, considering a variation of refractive index of the ITO active layer from 1.96+0.002i to 1.04+0.28i, respectively. (e) Experimentally measured transmission as function of bias voltage (-4V to 4V) of a 5 (red, rhombus) and 20 (blue, circles) $\mu$m ITO based EA modulator.}
\end{figure}
\begin{table}[h!]
\begin{tabular}{|l|l|l|l|l|l|l|}
\hline
Material                               & \multicolumn{1}{c|}{\begin{tabular}[c]{@{}c@{}}Device\\ type\end{tabular}} & \multicolumn{1}{c|}{\begin{tabular}[c]{@{}c@{}}IL \\ {[}dB{]}\end{tabular}} & \multicolumn{1}{c|}{\begin{tabular}[c]{@{}c@{}}Linear Footprint \\ {[}$\mu$m{]}\end{tabular}} & \multicolumn{1}{c|}{\begin{tabular}[c]{@{}c@{}}Speed \\ {[}Gbit/s{]}\end{tabular}} & \multicolumn{1}{c|}{\begin{tabular}[c]{@{}c@{}}Energy \\ {[}fJ/bit{]}\end{tabular}} & \multicolumn{1}{c|}{\begin{tabular}[c]{@{}c@{}}ER \\ {[}dB{]}\end{tabular}} \\ \hline
LiNbO$_3$\cite{wang_nanophotonic_2018}                    & MZM                                                                        & 2                                & 7000                                                                               & 60                                                                                 & 1700                                                                                & 8                                                                           \\ \hline
Si\cite{patel_frequency_2018}                        & MZM                                                                        & 10                               & 10500                                                                              & 4                                                                              & -                                                                                   & -                                                                           \\ \hline
Si\cite{streshinsky_low_2013}   & MZM                                                                        & 2                              & 3000                                                                               & 50                                                                                 & 450                                                                                 & 3                                                                         \\ \hline
Si\cite{li_40_2012}     & MRR                                                                        & 3                                & 10                                                                                 & 40                                                                                 & 80                                                                                  & 7                                                                           \\ \hline
Si\cite{pantouvaki_56gb/s_2015}           & MRR                                                                        & 3                                & 10                                                                                 & 56                                                                                 & 45                                                                                  & 4                                                                           \\ \hline
Si\cite{timurdogan_ultralow_2014} & MRR                                                                        & 1                                & 5                                                                                  & 40                                                                                 & 4                                                                                & 8                                                                           \\ \hline
\textbf{This work}                     & \textbf{EAM}                                                               & \textbf{1}                       & \textbf{5}                                                                         & \textbf{25}\cite{kim_silicon_2016}                                                        & \textbf{60}                                                                         & \textbf{5}                                                                  \\ \hline
\end{tabular}
\label{tab:table1}
\caption{Comparison to state of the art Silicon and LiNbO$_3$ based traveling-wave modulators, currently employed}
\end{table}

\begin{figure}[h!]
\includegraphics[width=.45\textwidth]{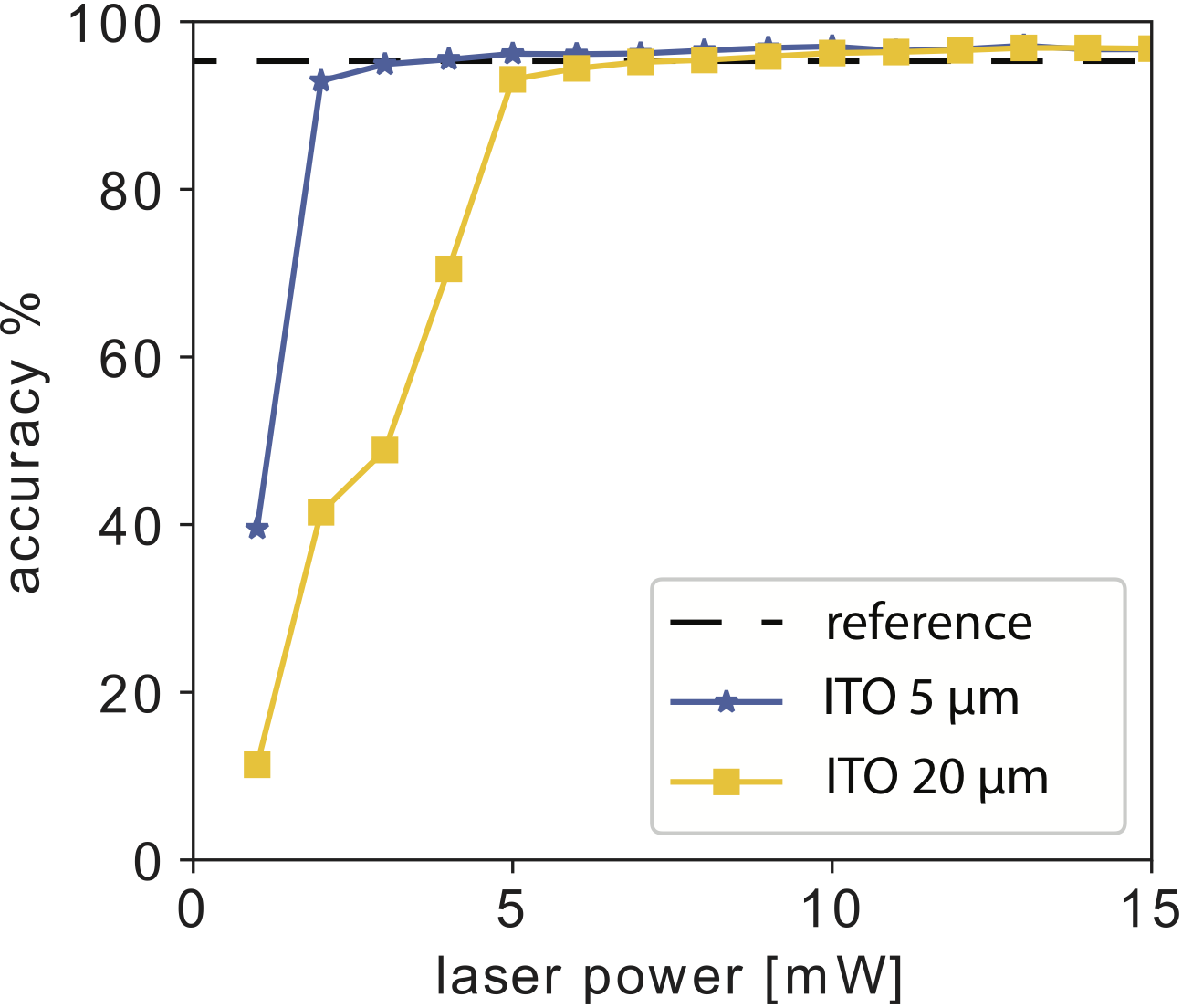}
\caption{\label{fig:ito_experiment_fit_nn_results}
Simulated MNIST accuracy for activation functions fit to experimental voltage absorption data from two ITO modulators, $5 \mu m$ and $20 \mu m$ \cite{sorger_ultra-compact_2012}. A 10 degree clipped polynomial (inset) was used to fit the transmittance vs. voltage data from each modulator. Laser power was swept from 1 mW to 15 mW over the fully connected neural network of two layers each with 100 nodes with both modulator types achieving greater than 90\% accuracy for laser power $ \geq $ 5 mW.
}
\end{figure}

\subsection{\label{sec:device}Experimental characterization of an ITO based Electro-Absorption Modulator Device}
Using the same approach presented in \cite{sorger_ultra-compact_2012}, we fabricate two different EA modulators with different length (5 and 20 $\mu$m). The device configuration consists of a Si waveguide (800 nm $\times$ 340 nm) and stack, placed on top, which comprises of 10 nm ITO layer, gate oxide (SiO$_2$, $t_{\text{ox}}$=20 nm) and a metallic gold pad. The stack was fabricated using electron-beam lithography for defining the pattern, electron-beam evaporation and lift off processing. 
In this capacitor configuration, according to the potential applied, we are able to modulate the carrier density of the ITO film, thus tuning the portion of the electric field absorbed by the thin layer. A 1550 nm TM mode travelling in the waveguide is subjected to a shift in the mode profile due to the presence of the plasmonic stack, and accordingly enhances the mode overlap with the active ITO layer. Increasing the carrier concentration level with active electro-static gating enhances the modulation effect due to the increased free carrier absorption of the optical mode dynamics as previously discussed.
The performances of the characterized electro-absorption modulators display extinction ratios of 5 dB and more than 20 dB for corresponding device lengths of 5 and 20 $\mu$m, respectively; when a voltage bias between the ITO and the metal contact is applied. The insertion losses are relatively low, and for a device of 5 $\mu$m are approximately 1 dB.
Going beyond the comparison between the proposed device and free carrier absorption in Si modulators, in Table \ref{tab:table1} we relate our ITO based EAM to the state-of-the-art Silicon and LiNbO$_3$ based traveling-wave modulators, currently employed in integrated photonic circuits.
As can therefore be inferred, the proposed modulator compares also very well with other regularly employed Si and Lithium Niobate state of the art EO modulators. The proposed modulator reaches striking modulation range considering the limited footprint. Moreover, it does not necessitate cavity feedback (eg. micro-ring resonators) and has a broadband response.

Proceeding further with our analysis, we introduce the experimental EA modulator performance as component of the Nonlinear activation function in the MNIST classifier of handwritten digits Fig. \ref{fig:epsart} as previously done for our theoretical study. We fit a 10 degree polynomial with saturation to the experimental data (Fig. \ref{fig:ito_experiment_fit_nn_results}). In this case with the greater performance of the plasmonic mode we modeled an activation function with the modulators capacitively coupled to the photodiodes. In capacitive coupling, the signal is integrated over one clock period and then allowed to modulate for the second clock period. This results in an 5 GHz effective clock rate at a 10 GHz frequency. The results show greater performance of the 5 $\mu$m ITO modulator over the 20 $\mu$m ITO modulator for laser power less than 5 mW, while both modulators asymptotically converge to an accuracy greater of 97\% as laser power is increased to 15 mW.

\section{\label{sec:conclusions}Conclusions}
In conclusion, in this work we described the free carrier absorption dynamics in ITO. The model of the ITO absorption mechanism is then incorporated as NL activation function of a neural network which comprises 2 layer with 100 nodes each. After an accurate multi-parameter optimization, the proposed noise-material capacitor-circuit model shows that the ITO based modulator used as component of the NL activation mechanism can provide high accuracy (up to 97\%) in the inference results in neural networks, which implements handwritten classification prediction tasks, with a throughput of of 5 GHz,
%\fxwarning{VS: pls check, above we used 5Ghz} 
characterized by low latency and a power budget in the order of less than 10 watts. For the optical platform only, in a trained network, the consumption of one electro-absorption module during the inference phase is assumed approximately to be 60 fJ/bit ( $1/2 CV^2$). At the system level, neglecting I/O and laser power, this translates into 12 pJ/bit ($200 \text{NL AF} \times60$fJ/bit) consumption for the linear activation functions. %\fxwarning{VS: could we compute a MAC/J for this? But it may depend on the #of nodes feed in the network. Well we use 200 here and I assume the laser power in Fig. 3 and 4 are for all 200 nodes.} 
Moreover, we incorporate in our photonic model of tensor-flow, experimental data of a fabricated ITO based EA modulator, showing even higher performance than the ones predicted by the model, which doesn't account for device asymmetries and material nonidealities.    
In our vision, these results make ITO a very competitive material for electro-optic neural network, particularly for implementing nonlinear activation function for low latency and power demanding applications such as communications and LiDAR.

\begin{acknowledgments}
V.S., T.E., and P.P. are supported via the E2CDA program: V.S., T.E., and P.P. under NSF grant number 1740262 and V.S. and T.E. under SRC nCORE grant number 1740235 and P.P. under SRC nCORE grant number 2018-NC-2763-A.
\end{acknowledgments}


\begin{thebibliography}{0}%
\makeatletter
\providecommand \@ifxundefined [1]{%
 \@ifx{#1\undefined}
}%
\providecommand \@ifnum [1]{%
 \ifnum #1\expandafter \@firstoftwo
 \else \expandafter \@secondoftwo
 \fi
}%
\providecommand \@ifx [1]{%
 \ifx #1\expandafter \@firstoftwo
 \else \expandafter \@secondoftwo
 \fi
}%
\providecommand \natexlab [1]{#1}%
\providecommand \enquote  [1]{``#1''}%
\providecommand \bibnamefont  [1]{#1}%
\providecommand \bibfnamefont [1]{#1}%
\providecommand \citenamefont [1]{#1}%
\providecommand \href@noop [0]{\@secondoftwo}%
\providecommand \href [0]{\begingroup \@sanitize@url \@href}%
\providecommand \@href[1]{\@@startlink{#1}\@@href}%
\providecommand \@@href[1]{\endgroup#1\@@endlink}%
\providecommand \@sanitize@url [0]{\catcode `\\12\catcode `\$12\catcode
  `\&12\catcode `\#12\catcode `\^12\catcode `\_12\catcode `\%12\relax}%
\providecommand \@@startlink[1]{}%
\providecommand \@@endlink[0]{}%
\providecommand \url  [0]{\begingroup\@sanitize@url \@url }%
\providecommand \@url [1]{\endgroup\@href {#1}{\urlprefix }}%
\providecommand \urlprefix  [0]{URL }%
\providecommand \Eprint [0]{\href }%
\providecommand \doibase [0]{http://dx.doi.org/}%
\providecommand \selectlanguage [0]{\@gobble}%
\providecommand \bibinfo  [0]{\@secondoftwo}%
\providecommand \bibfield  [0]{\@secondoftwo}%
\providecommand \translation [1]{[#1]}%
\providecommand \BibitemOpen [0]{}%
\providecommand \bibitemStop [0]{}%
\providecommand \bibitemNoStop [0]{.\EOS\space}%
\providecommand \EOS [0]{\spacefactor3000\relax}%
\providecommand \BibitemShut  [1]{\csname bibitem#1\endcsname}%
\let\auto@bib@innerbib\@empty
%</preamble>
\end{thebibliography}%



\begin{thebibliography}{75}%
\makeatletter
\providecommand \@ifxundefined [1]{%
 \@ifx{#1\undefined}
}%
\providecommand \@ifnum [1]{%
 \ifnum #1\expandafter \@firstoftwo
 \else \expandafter \@secondoftwo
 \fi
}%
\providecommand \@ifx [1]{%
 \ifx #1\expandafter \@firstoftwo
 \else \expandafter \@secondoftwo
 \fi
}%
\providecommand \natexlab [1]{#1}%
\providecommand \enquote  [1]{``#1''}%
\providecommand \bibnamefont  [1]{#1}%
\providecommand \bibfnamefont [1]{#1}%
\providecommand \citenamefont [1]{#1}%
\providecommand \href@noop [0]{\@secondoftwo}%
\providecommand \href [0]{\begingroup \@sanitize@url \@href}%
\providecommand \@href[1]{\@@startlink{#1}\@@href}%
\providecommand \@@href[1]{\endgroup#1\@@endlink}%
\providecommand \@sanitize@url [0]{\catcode `\\12\catcode `\$12\catcode
  `\&12\catcode `\#12\catcode `\^12\catcode `\_12\catcode `\%12\relax}%
\providecommand \@@startlink[1]{}%
\providecommand \@@endlink[0]{}%
\providecommand \url  [0]{\begingroup\@sanitize@url \@url }%
\providecommand \@url [1]{\endgroup\@href {#1}{\urlprefix }}%
\providecommand \urlprefix  [0]{URL }%
\providecommand \Eprint [0]{\href }%
\providecommand \doibase [0]{http://dx.doi.org/}%
\providecommand \selectlanguage [0]{\@gobble}%
\providecommand \bibinfo  [0]{\@secondoftwo}%
\providecommand \bibfield  [0]{\@secondoftwo}%
\providecommand \translation [1]{[#1]}%
\providecommand \BibitemOpen [0]{}%
\providecommand \bibitemStop [0]{}%
\providecommand \bibitemNoStop [0]{.\EOS\space}%
\providecommand \EOS [0]{\spacefactor3000\relax}%
\providecommand \BibitemShut  [1]{\csname bibitem#1\endcsname}%
\let\auto@bib@innerbib\@empty
%</preamble>
\bibitem [{\citenamefont {Prucnal}\ and\ \citenamefont
  {Shastri}(2017)}]{prucnal_neuromorphic_2017}%
  \BibitemOpen
  \bibinfo {editor} {\bibfnamefont {P.~R.}\ \bibnamefont {Prucnal}}\ and\
  \bibinfo {editor} {\bibfnamefont {B.~J.}\ \bibnamefont {Shastri}},\ eds.,\
  \href@noop {} {{\selectlanguage {English}\emph {\bibinfo {title}
  {Neuromorphic {Photonics}}}}},\ \bibinfo {edition} {1st}\ ed.\ (\bibinfo
  {publisher} {CRC Press},\ \bibinfo {address} {Boca Raton, FL},\ \bibinfo
  {year} {2017})\BibitemShut {NoStop}%
\bibitem [{\citenamefont {Shastri}\ \emph {et~al.}(2018)\citenamefont
  {Shastri}, \citenamefont {Tait}, \citenamefont {de~Lima}, \citenamefont
  {Nahmias}, \citenamefont {Peng},\ and\ \citenamefont
  {Prucnal}}]{shastri_principles_2018}%
  \BibitemOpen
  \bibfield  {author} {\bibinfo {author} {\bibfnamefont {B.~J.}\ \bibnamefont
  {Shastri}}, \bibinfo {author} {\bibfnamefont {A.~N.}\ \bibnamefont {Tait}},
  \bibinfo {author} {\bibfnamefont {T.~F.}\ \bibnamefont {de~Lima}}, \bibinfo
  {author} {\bibfnamefont {M.~A.}\ \bibnamefont {Nahmias}}, \bibinfo {author}
  {\bibfnamefont {H.-T.}\ \bibnamefont {Peng}}, \ and\ \bibinfo {author}
  {\bibfnamefont {P.~R.}\ \bibnamefont {Prucnal}},\ }\bibfield  {title}
  {\enquote {\bibinfo {title} {Principles of neuromorphic photonics},}\ }\href
  {\doibase 10.1007/978-3-642-27737-5_702-1} {\bibfield  {journal} {\bibinfo
  {journal} {arXiv:1801.00016 [physics]}\ ,\ \bibinfo {pages} {1--37}}
  (\bibinfo {year} {2018})},\ \bibinfo {note} {arXiv: 1801.00016}\BibitemShut
  {NoStop}%
\bibitem [{\citenamefont {Shen}\ \emph {et~al.}(2017)\citenamefont {Shen},
  \citenamefont {Harris}, \citenamefont {Skirlo}, \citenamefont {Prabhu},
  \citenamefont {Baehr-Jones}, \citenamefont {Hochberg}, \citenamefont {Sun},
  \citenamefont {Zhao}, \citenamefont {Larochelle}, \citenamefont {Englund},\
  and\ \citenamefont {Soljačić}}]{shen_deep_2017}%
  \BibitemOpen
  \bibfield  {author} {\bibinfo {author} {\bibfnamefont {Y.}~\bibnamefont
  {Shen}}, \bibinfo {author} {\bibfnamefont {N.~C.}\ \bibnamefont {Harris}},
  \bibinfo {author} {\bibfnamefont {S.}~\bibnamefont {Skirlo}}, \bibinfo
  {author} {\bibfnamefont {M.}~\bibnamefont {Prabhu}}, \bibinfo {author}
  {\bibfnamefont {T.}~\bibnamefont {Baehr-Jones}}, \bibinfo {author}
  {\bibfnamefont {M.}~\bibnamefont {Hochberg}}, \bibinfo {author}
  {\bibfnamefont {X.}~\bibnamefont {Sun}}, \bibinfo {author} {\bibfnamefont
  {S.}~\bibnamefont {Zhao}}, \bibinfo {author} {\bibfnamefont {H.}~\bibnamefont
  {Larochelle}}, \bibinfo {author} {\bibfnamefont {D.}~\bibnamefont {Englund}},
  \ and\ \bibinfo {author} {\bibfnamefont {M.}~\bibnamefont {Soljačić}},\
  }\bibfield  {title} {\enquote {\bibinfo {title} {Deep learning with coherent
  nanophotonic circuits},}\ }\href {\doibase 10.1038/nphoton.2017.93}
  {\bibfield  {journal} {\bibinfo  {journal} {Nature Photonics}\ }\textbf
  {\bibinfo {volume} {11}},\ \bibinfo {pages} {441--446} (\bibinfo {year}
  {2017})}\BibitemShut {NoStop}%
\bibitem [{\citenamefont {Heni}\ \emph {et~al.}(2019)\citenamefont {Heni},
  \citenamefont {Fedoryshyn}, \citenamefont {Baeuerle}, \citenamefont {Josten},
  \citenamefont {Hoessbacher}, \citenamefont {Messner}, \citenamefont
  {Haffner}, \citenamefont {Watanabe}, \citenamefont {Salamin}, \citenamefont
  {Koch}, \citenamefont {Elder}, \citenamefont {Dalton},\ and\ \citenamefont
  {Leuthold}}]{heni_plasmonic_2019}%
  \BibitemOpen
  \bibfield  {author} {\bibinfo {author} {\bibfnamefont {W.}~\bibnamefont
  {Heni}}, \bibinfo {author} {\bibfnamefont {Y.}~\bibnamefont {Fedoryshyn}},
  \bibinfo {author} {\bibfnamefont {B.}~\bibnamefont {Baeuerle}}, \bibinfo
  {author} {\bibfnamefont {A.}~\bibnamefont {Josten}}, \bibinfo {author}
  {\bibfnamefont {C.~B.}\ \bibnamefont {Hoessbacher}}, \bibinfo {author}
  {\bibfnamefont {A.}~\bibnamefont {Messner}}, \bibinfo {author} {\bibfnamefont
  {C.}~\bibnamefont {Haffner}}, \bibinfo {author} {\bibfnamefont
  {T.}~\bibnamefont {Watanabe}}, \bibinfo {author} {\bibfnamefont
  {Y.}~\bibnamefont {Salamin}}, \bibinfo {author} {\bibfnamefont
  {U.}~\bibnamefont {Koch}}, \bibinfo {author} {\bibfnamefont {D.~L.}\
  \bibnamefont {Elder}}, \bibinfo {author} {\bibfnamefont {L.~R.}\ \bibnamefont
  {Dalton}}, \ and\ \bibinfo {author} {\bibfnamefont {J.}~\bibnamefont
  {Leuthold}},\ }\bibfield  {title} {\enquote {\bibinfo {title} {Plasmonic {IQ}
  modulators with attojoule per bit electrical energy consumption},}\ }\href
  {\doibase 10.1038/s41467-019-09724-7} {\bibfield  {journal} {\bibinfo
  {journal} {Nature Communications}\ }\textbf {\bibinfo {volume} {10}},\
  \bibinfo {pages} {1694} (\bibinfo {year} {2019})}\BibitemShut {NoStop}%
\bibitem [{\citenamefont {Amin}\ \emph
  {et~al.}(2018{\natexlab{a}})\citenamefont {Amin}, \citenamefont {Ma},
  \citenamefont {Maiti}, \citenamefont {Khan}, \citenamefont {Khurgin},
  \citenamefont {Dalir},\ and\ \citenamefont
  {Sorger}}]{amin_attojoule-efficient_2018}%
  \BibitemOpen
  \bibfield  {author} {\bibinfo {author} {\bibfnamefont {R.}~\bibnamefont
  {Amin}}, \bibinfo {author} {\bibfnamefont {Z.}~\bibnamefont {Ma}}, \bibinfo
  {author} {\bibfnamefont {R.}~\bibnamefont {Maiti}}, \bibinfo {author}
  {\bibfnamefont {S.}~\bibnamefont {Khan}}, \bibinfo {author} {\bibfnamefont
  {J.~B.}\ \bibnamefont {Khurgin}}, \bibinfo {author} {\bibfnamefont
  {H.}~\bibnamefont {Dalir}}, \ and\ \bibinfo {author} {\bibfnamefont {V.~J.}\
  \bibnamefont {Sorger}},\ }\bibfield  {title} {\enquote {\bibinfo {title}
  {Attojoule-efficient graphene optical modulators},}\ }\href {\doibase
  10.1364/AO.57.00D130} {\bibfield  {journal} {\bibinfo  {journal} {Applied
  Optics}\ }\textbf {\bibinfo {volume} {57}},\ \bibinfo {pages} {D130--D140}
  (\bibinfo {year} {2018}{\natexlab{a}})}\BibitemShut {NoStop}%
\bibitem [{\citenamefont {Nozaki}\ \emph {et~al.}(2017)\citenamefont {Nozaki},
  \citenamefont {Shakoor}, \citenamefont {Matsuo}, \citenamefont {Fujii},
  \citenamefont {Takeda}, \citenamefont {Shinya}, \citenamefont {Kuramochi},\
  and\ \citenamefont {Notomi}}]{nozaki_ultralow-energy_2017}%
  \BibitemOpen
  \bibfield  {author} {\bibinfo {author} {\bibfnamefont {K.}~\bibnamefont
  {Nozaki}}, \bibinfo {author} {\bibfnamefont {A.}~\bibnamefont {Shakoor}},
  \bibinfo {author} {\bibfnamefont {S.}~\bibnamefont {Matsuo}}, \bibinfo
  {author} {\bibfnamefont {T.}~\bibnamefont {Fujii}}, \bibinfo {author}
  {\bibfnamefont {K.}~\bibnamefont {Takeda}}, \bibinfo {author} {\bibfnamefont
  {A.}~\bibnamefont {Shinya}}, \bibinfo {author} {\bibfnamefont
  {E.}~\bibnamefont {Kuramochi}}, \ and\ \bibinfo {author} {\bibfnamefont
  {M.}~\bibnamefont {Notomi}},\ }\bibfield  {title} {\enquote {\bibinfo {title}
  {Ultralow-energy electro-absorption modulator consisting of
  {InGaAsP}-embedded photonic-crystal waveguide},}\ }\href {\doibase
  10.1063/1.4980036} {\bibfield  {journal} {\bibinfo  {journal} {APL
  Photonics}\ }\textbf {\bibinfo {volume} {2}},\ \bibinfo {pages} {056105}
  (\bibinfo {year} {2017})}\BibitemShut {NoStop}%
\bibitem [{\citenamefont {Chiu}(2015)}]{chiu_design_2015}%
  \BibitemOpen
  \bibfield  {author} {\bibinfo {author} {\bibfnamefont {Y.}~\bibnamefont
  {Chiu}},\ }\bibfield  {title} {\enquote {\bibinfo {title} {Design and
  fabrication of optical electroabsorption modulator for high speed and high
  efficiency},}\ }in\ \href {\doibase 10.1109/ICOCN.2015.7203651} {\emph
  {\bibinfo {booktitle} {2015 14th {International} {Conference} on {Optical}
  {Communications} and {Networks} ({ICOCN})}}}\ (\bibinfo {year} {2015})\ pp.\
  \bibinfo {pages} {1--4}\BibitemShut {NoStop}%
\bibitem [{\citenamefont {Ferreira}\ \emph {et~al.}(2017)\citenamefont
  {Ferreira}, \citenamefont {Shastri}, \citenamefont {Tait}, \citenamefont
  {Nahmias},\ and\ \citenamefont {Prucnal}}]{ferreira_progress_2017}%
  \BibitemOpen
  \bibfield  {author} {\bibinfo {author} {\bibfnamefont {d.~L.~T.}\
  \bibnamefont {Ferreira}}, \bibinfo {author} {\bibfnamefont {B.~J.}\
  \bibnamefont {Shastri}}, \bibinfo {author} {\bibfnamefont {A.~N.}\
  \bibnamefont {Tait}}, \bibinfo {author} {\bibfnamefont {M.~A.}\ \bibnamefont
  {Nahmias}}, \ and\ \bibinfo {author} {\bibfnamefont {P.~R.}\ \bibnamefont
  {Prucnal}},\ }\bibfield  {title} {\enquote {\bibinfo {title} {Progress in
  neuromorphic photonics},}\ }\href {\doibase 10.1515/nanoph-2016-0139}
  {\bibfield  {journal} {\bibinfo  {journal} {Nanophotonics}\ }\textbf
  {\bibinfo {volume} {6}},\ \bibinfo {pages} {577--599} (\bibinfo {year}
  {2017})}\BibitemShut {NoStop}%
\bibitem [{\citenamefont {Brunner}\ \emph {et~al.}(2013)\citenamefont
  {Brunner}, \citenamefont {Soriano}, \citenamefont {Mirasso},\ and\
  \citenamefont {Fischer}}]{brunner_parallel_2013}%
  \BibitemOpen
  \bibfield  {author} {\bibinfo {author} {\bibfnamefont {D.}~\bibnamefont
  {Brunner}}, \bibinfo {author} {\bibfnamefont {M.~C.}\ \bibnamefont
  {Soriano}}, \bibinfo {author} {\bibfnamefont {C.~R.}\ \bibnamefont
  {Mirasso}}, \ and\ \bibinfo {author} {\bibfnamefont {I.}~\bibnamefont
  {Fischer}},\ }\bibfield  {title} {\enquote {\bibinfo {title} {Parallel
  photonic information processing at gigabyte per second data rates using
  transient states},}\ }\href {\doibase 10.1038/ncomms2368} {\bibfield
  {journal} {\bibinfo  {journal} {Nature Communications}\ }\textbf {\bibinfo
  {volume} {4}},\ \bibinfo {pages} {1364} (\bibinfo {year} {2013})}\BibitemShut
  {NoStop}%
\bibitem [{\citenamefont {Blumenthal}(2018)}]{blumenthal_integrated_2018}%
  \BibitemOpen
  \bibfield  {author} {\bibinfo {author} {\bibfnamefont {D.~J.}\ \bibnamefont
  {Blumenthal}},\ }\bibfield  {title} {\enquote {\bibinfo {title} {Integrated
  combs drive extreme data rates},}\ }\href {\doibase
  10.1038/s41566-018-0222-4} {\bibfield  {journal} {\bibinfo  {journal} {Nature
  Photonics}\ }\textbf {\bibinfo {volume} {12}},\ \bibinfo {pages} {447}
  (\bibinfo {year} {2018})}\BibitemShut {NoStop}%
\bibitem [{\citenamefont {Moscoso-Mártir}\ \emph {et~al.}(2017)\citenamefont
  {Moscoso-Mártir}, \citenamefont {Müller}, \citenamefont {Islamova},
  \citenamefont {Merget},\ and\ \citenamefont
  {Witzens}}]{moscoso-martir_calibrated_2017}%
  \BibitemOpen
  \bibfield  {author} {\bibinfo {author} {\bibfnamefont {A.}~\bibnamefont
  {Moscoso-Mártir}}, \bibinfo {author} {\bibfnamefont {J.}~\bibnamefont
  {Müller}}, \bibinfo {author} {\bibfnamefont {E.}~\bibnamefont {Islamova}},
  \bibinfo {author} {\bibfnamefont {F.}~\bibnamefont {Merget}}, \ and\ \bibinfo
  {author} {\bibfnamefont {J.}~\bibnamefont {Witzens}},\ }\bibfield  {title}
  {\enquote {\bibinfo {title} {Calibrated {Link} {Budget} of a {Silicon}
  {Photonics} {WDM} {Transceiver} with {SOA} and {Semiconductor}
  {Mode}-{Locked} {Laser}},}\ }\href {\doibase 10.1038/s41598-017-12023-0}
  {\bibfield  {journal} {\bibinfo  {journal} {Scientific Reports}\ }\textbf
  {\bibinfo {volume} {7}},\ \bibinfo {pages} {12004} (\bibinfo {year}
  {2017})}\BibitemShut {NoStop}%
\bibitem [{\citenamefont {Mehrabian}\ \emph {et~al.}(2018)\citenamefont
  {Mehrabian}, \citenamefont {Al-Kabani}, \citenamefont {Sorger},\ and\
  \citenamefont {El-Ghazawi}}]{mehrabian_pcnna:_2018}%
  \BibitemOpen
  \bibfield  {author} {\bibinfo {author} {\bibfnamefont {A.}~\bibnamefont
  {Mehrabian}}, \bibinfo {author} {\bibfnamefont {Y.}~\bibnamefont
  {Al-Kabani}}, \bibinfo {author} {\bibfnamefont {V.~J.}\ \bibnamefont
  {Sorger}}, \ and\ \bibinfo {author} {\bibfnamefont {T.}~\bibnamefont
  {El-Ghazawi}},\ }\bibfield  {title} {\enquote {\bibinfo {title} {{PCNNA}: {A}
  {Photonic} {Convolutional} {Neural} {Network} {Accelerator}},}\ }\href
  {https://arxiv.org/abs/1807.08792v1} {\bibfield  {journal} {\bibinfo
  {journal} {arXiv}\ } (\bibinfo {year} {2018})}\BibitemShut {NoStop}%
\bibitem [{\citenamefont {Tait}\ \emph {et~al.}(2014)\citenamefont {Tait},
  \citenamefont {Nahmias}, \citenamefont {Shastri},\ and\ \citenamefont
  {Prucnal}}]{tait_broadcast_2014}%
  \BibitemOpen
  \bibfield  {author} {\bibinfo {author} {\bibfnamefont {A.~N.}\ \bibnamefont
  {Tait}}, \bibinfo {author} {\bibfnamefont {M.~A.}\ \bibnamefont {Nahmias}},
  \bibinfo {author} {\bibfnamefont {B.~J.}\ \bibnamefont {Shastri}}, \ and\
  \bibinfo {author} {\bibfnamefont {P.~R.}\ \bibnamefont {Prucnal}},\
  }\bibfield  {title} {\enquote {\bibinfo {title} {Broadcast and {Weight}: {An}
  {Integrated} {Network} {For} {Scalable} {Photonic} {Spike} {Processing}},}\
  }\href {https://www.osapublishing.org/jlt/abstract.cfm?uri=jlt-32-21-3427}
  {\bibfield  {journal} {\bibinfo  {journal} {Journal of Lightwave Technology}\
  }\textbf {\bibinfo {volume} {32}},\ \bibinfo {pages} {3427--3439} (\bibinfo
  {year} {2014})}\BibitemShut {NoStop}%
\bibitem [{\citenamefont {George}\ \emph {et~al.}(2019)\citenamefont {George},
  \citenamefont {Mehrabian}, \citenamefont {Amin}, \citenamefont {Meng},
  \citenamefont {Lima}, \citenamefont {Tait}, \citenamefont {Shastri},
  \citenamefont {El-Ghazawi}, \citenamefont {Prucnal},\ and\ \citenamefont
  {Sorger}}]{george_neuromorphic_2019}%
  \BibitemOpen
  \bibfield  {author} {\bibinfo {author} {\bibfnamefont {J.~K.}\ \bibnamefont
  {George}}, \bibinfo {author} {\bibfnamefont {A.}~\bibnamefont {Mehrabian}},
  \bibinfo {author} {\bibfnamefont {R.}~\bibnamefont {Amin}}, \bibinfo {author}
  {\bibfnamefont {J.}~\bibnamefont {Meng}}, \bibinfo {author} {\bibfnamefont
  {T.~F.~d.}\ \bibnamefont {Lima}}, \bibinfo {author} {\bibfnamefont {A.~N.}\
  \bibnamefont {Tait}}, \bibinfo {author} {\bibfnamefont {B.~J.}\ \bibnamefont
  {Shastri}}, \bibinfo {author} {\bibfnamefont {T.}~\bibnamefont {El-Ghazawi}},
  \bibinfo {author} {\bibfnamefont {P.~R.}\ \bibnamefont {Prucnal}}, \ and\
  \bibinfo {author} {\bibfnamefont {V.~J.}\ \bibnamefont {Sorger}},\ }\bibfield
   {title} {\enquote {\bibinfo {title} {Neuromorphic photonics with
  electro-absorption modulators},}\ }\href {\doibase 10.1364/OE.27.005181}
  {\bibfield  {journal} {\bibinfo  {journal} {Optics Express}\ }\textbf
  {\bibinfo {volume} {27}},\ \bibinfo {pages} {5181--5191} (\bibinfo {year}
  {2019})}\BibitemShut {NoStop}%
\bibitem [{\citenamefont {Miscuglio}\ \emph {et~al.}(2018)\citenamefont
  {Miscuglio}, \citenamefont {Mehrabian}, \citenamefont {Hu}, \citenamefont
  {Azzam}, \citenamefont {George}, \citenamefont {Kildishev}, \citenamefont
  {Pelton},\ and\ \citenamefont {Sorger}}]{miscuglio_all-optical_2018}%
  \BibitemOpen
  \bibfield  {author} {\bibinfo {author} {\bibfnamefont {M.}~\bibnamefont
  {Miscuglio}}, \bibinfo {author} {\bibfnamefont {A.}~\bibnamefont
  {Mehrabian}}, \bibinfo {author} {\bibfnamefont {Z.}~\bibnamefont {Hu}},
  \bibinfo {author} {\bibfnamefont {S.~I.}\ \bibnamefont {Azzam}}, \bibinfo
  {author} {\bibfnamefont {J.}~\bibnamefont {George}}, \bibinfo {author}
  {\bibfnamefont {A.~V.}\ \bibnamefont {Kildishev}}, \bibinfo {author}
  {\bibfnamefont {M.}~\bibnamefont {Pelton}}, \ and\ \bibinfo {author}
  {\bibfnamefont {V.~J.}\ \bibnamefont {Sorger}},\ }\bibfield  {title}
  {\enquote {\bibinfo {title} {All-optical nonlinear activation function for
  photonic neural networks [{Invited}]},}\ }\href {\doibase
  10.1364/OME.8.003851} {\bibfield  {journal} {\bibinfo  {journal} {Optical
  Materials Express}\ }\textbf {\bibinfo {volume} {8}},\ \bibinfo {pages}
  {3851--3863} (\bibinfo {year} {2018})}\BibitemShut {NoStop}%
\bibitem [{\citenamefont {Dejonckheere}\ \emph {et~al.}(2014)\citenamefont
  {Dejonckheere}, \citenamefont {Duport}, \citenamefont {Smerieri},
  \citenamefont {Fang}, \citenamefont {Oudar}, \citenamefont {Haelterman},\
  and\ \citenamefont {Massar}}]{dejonckheere_all-optical_2014}%
  \BibitemOpen
  \bibfield  {author} {\bibinfo {author} {\bibfnamefont {A.}~\bibnamefont
  {Dejonckheere}}, \bibinfo {author} {\bibfnamefont {F.}~\bibnamefont
  {Duport}}, \bibinfo {author} {\bibfnamefont {A.}~\bibnamefont {Smerieri}},
  \bibinfo {author} {\bibfnamefont {L.}~\bibnamefont {Fang}}, \bibinfo {author}
  {\bibfnamefont {J.-L.}\ \bibnamefont {Oudar}}, \bibinfo {author}
  {\bibfnamefont {M.}~\bibnamefont {Haelterman}}, \ and\ \bibinfo {author}
  {\bibfnamefont {S.}~\bibnamefont {Massar}},\ }\bibfield  {title} {\enquote
  {\bibinfo {title} {All-optical reservoir computer based on saturation of
  absorption},}\ }\href {\doibase 10.1364/OE.22.010868} {\bibfield  {journal}
  {\bibinfo  {journal} {Optics Express}\ }\textbf {\bibinfo {volume} {22}},\
  \bibinfo {pages} {10868} (\bibinfo {year} {2014})}\BibitemShut {NoStop}%
\bibitem [{\citenamefont {Ma}\ \emph {et~al.}(2017)\citenamefont {Ma},
  \citenamefont {Shastri}, \citenamefont {de~Lima}, \citenamefont {Tait},
  \citenamefont {Nahmias},\ and\ \citenamefont
  {Prucnal}}]{ma_all-optical_2017}%
  \BibitemOpen
  \bibfield  {author} {\bibinfo {author} {\bibfnamefont {P.~Y.}\ \bibnamefont
  {Ma}}, \bibinfo {author} {\bibfnamefont {B.~J.}\ \bibnamefont {Shastri}},
  \bibinfo {author} {\bibfnamefont {T.~F.}\ \bibnamefont {de~Lima}}, \bibinfo
  {author} {\bibfnamefont {A.~N.}\ \bibnamefont {Tait}}, \bibinfo {author}
  {\bibfnamefont {M.~A.}\ \bibnamefont {Nahmias}}, \ and\ \bibinfo {author}
  {\bibfnamefont {P.~R.}\ \bibnamefont {Prucnal}},\ }\bibfield  {title}
  {\enquote {\bibinfo {title} {All-optical digital-to-spike conversion using a
  graphene excitable laser},}\ }\href {\doibase 10.1364/OE.25.033504}
  {\bibfield  {journal} {\bibinfo  {journal} {Optics Express}\ }\textbf
  {\bibinfo {volume} {25}},\ \bibinfo {pages} {33504} (\bibinfo {year}
  {2017})}\BibitemShut {NoStop}%
\bibitem [{\citenamefont {Shastri}\ \emph {et~al.}(2016)\citenamefont
  {Shastri}, \citenamefont {Nahmias}, \citenamefont {Tait}, \citenamefont
  {Rodriguez}, \citenamefont {Wu},\ and\ \citenamefont
  {Prucnal}}]{shastri_spike_2016}%
  \BibitemOpen
  \bibfield  {author} {\bibinfo {author} {\bibfnamefont {B.~J.}\ \bibnamefont
  {Shastri}}, \bibinfo {author} {\bibfnamefont {M.~A.}\ \bibnamefont
  {Nahmias}}, \bibinfo {author} {\bibfnamefont {A.~N.}\ \bibnamefont {Tait}},
  \bibinfo {author} {\bibfnamefont {A.~W.}\ \bibnamefont {Rodriguez}}, \bibinfo
  {author} {\bibfnamefont {B.}~\bibnamefont {Wu}}, \ and\ \bibinfo {author}
  {\bibfnamefont {P.~R.}\ \bibnamefont {Prucnal}},\ }\bibfield  {title}
  {\enquote {\bibinfo {title} {Spike processing with a graphene excitable
  laser},}\ }\href {\doibase 10.1038/srep19126} {\bibfield  {journal} {\bibinfo
   {journal} {Scientific Reports}\ }\textbf {\bibinfo {volume} {6}} (\bibinfo
  {year} {2016}),\ 10.1038/srep19126}\BibitemShut {NoStop}%
\bibitem [{\citenamefont {Peng}\ \emph {et~al.}(2018)\citenamefont {Peng},
  \citenamefont {Nahmias}, \citenamefont {Lima}, \citenamefont {Tait},\ and\
  \citenamefont {Shastri}}]{peng_neuromorphic_2018}%
  \BibitemOpen
  \bibfield  {author} {\bibinfo {author} {\bibfnamefont {H.}~\bibnamefont
  {Peng}}, \bibinfo {author} {\bibfnamefont {M.~A.}\ \bibnamefont {Nahmias}},
  \bibinfo {author} {\bibfnamefont {T.~F.~d.}\ \bibnamefont {Lima}}, \bibinfo
  {author} {\bibfnamefont {A.~N.}\ \bibnamefont {Tait}}, \ and\ \bibinfo
  {author} {\bibfnamefont {B.~J.}\ \bibnamefont {Shastri}},\ }\bibfield
  {title} {\enquote {\bibinfo {title} {Neuromorphic {Photonic} {Integrated}
  {Circuits}},}\ }\href {\doibase 10.1109/JSTQE.2018.2840448} {\bibfield
  {journal} {\bibinfo  {journal} {IEEE Journal of Selected Topics in Quantum
  Electronics}\ }\textbf {\bibinfo {volume} {24}},\ \bibinfo {pages} {1--15}
  (\bibinfo {year} {2018})}\BibitemShut {NoStop}%
\bibitem [{\citenamefont {Mesaritakis}\ \emph {et~al.}(2016)\citenamefont
  {Mesaritakis}, \citenamefont {Kapsalis}, \citenamefont {Bogris},\ and\
  \citenamefont {Syvridis}}]{mesaritakis_artificial_2016}%
  \BibitemOpen
  \bibfield  {author} {\bibinfo {author} {\bibfnamefont {C.}~\bibnamefont
  {Mesaritakis}}, \bibinfo {author} {\bibfnamefont {A.}~\bibnamefont
  {Kapsalis}}, \bibinfo {author} {\bibfnamefont {A.}~\bibnamefont {Bogris}}, \
  and\ \bibinfo {author} {\bibfnamefont {D.}~\bibnamefont {Syvridis}},\
  }\bibfield  {title} {\enquote {\bibinfo {title} {Artificial {Neuron} {Based}
  on {Integrated} {Semiconductor} {Quantum} {Dot} {Mode}-{Locked} {Lasers}},}\
  }\href {\doibase 10.1038/srep39317} {\bibfield  {journal} {\bibinfo
  {journal} {Scientific Reports}\ }\textbf {\bibinfo {volume} {6}},\ \bibinfo
  {pages} {39317} (\bibinfo {year} {2016})}\BibitemShut {NoStop}%
\bibitem [{\citenamefont {Alexander}\ \emph {et~al.}(2013)\citenamefont
  {Alexander}, \citenamefont {Vaerenbergh}, \citenamefont {Fiers},
  \citenamefont {Mechet}, \citenamefont {Dambre},\ and\ \citenamefont
  {Bienstman}}]{alexander_excitability_2013}%
  \BibitemOpen
  \bibfield  {author} {\bibinfo {author} {\bibfnamefont {K.}~\bibnamefont
  {Alexander}}, \bibinfo {author} {\bibfnamefont {T.~V.}\ \bibnamefont
  {Vaerenbergh}}, \bibinfo {author} {\bibfnamefont {M.}~\bibnamefont {Fiers}},
  \bibinfo {author} {\bibfnamefont {P.}~\bibnamefont {Mechet}}, \bibinfo
  {author} {\bibfnamefont {J.}~\bibnamefont {Dambre}}, \ and\ \bibinfo {author}
  {\bibfnamefont {P.}~\bibnamefont {Bienstman}},\ }\bibfield  {title} {\enquote
  {\bibinfo {title} {Excitability in optically injected microdisk lasers with
  phase controlled excitatory and inhibitory response},}\ }\href {\doibase
  10.1364/OE.21.026182} {\bibfield  {journal} {\bibinfo  {journal} {Optics
  Express}\ }\textbf {\bibinfo {volume} {21}},\ \bibinfo {pages} {26182--26191}
  (\bibinfo {year} {2013})}\BibitemShut {NoStop}%
\bibitem [{\citenamefont {Coarer}\ \emph {et~al.}(2018)\citenamefont {Coarer},
  \citenamefont {Sciamanna}, \citenamefont {Katumba}, \citenamefont
  {Freiberger}, \citenamefont {Dambre}, \citenamefont {Bienstman},\ and\
  \citenamefont {Rontani}}]{coarer_all-optical_2018}%
  \BibitemOpen
  \bibfield  {author} {\bibinfo {author} {\bibfnamefont {F.~D.}\ \bibnamefont
  {Coarer}}, \bibinfo {author} {\bibfnamefont {M.}~\bibnamefont {Sciamanna}},
  \bibinfo {author} {\bibfnamefont {A.}~\bibnamefont {Katumba}}, \bibinfo
  {author} {\bibfnamefont {M.}~\bibnamefont {Freiberger}}, \bibinfo {author}
  {\bibfnamefont {J.}~\bibnamefont {Dambre}}, \bibinfo {author} {\bibfnamefont
  {P.}~\bibnamefont {Bienstman}}, \ and\ \bibinfo {author} {\bibfnamefont
  {D.}~\bibnamefont {Rontani}},\ }\bibfield  {title} {\enquote {\bibinfo
  {title} {All-{Optical} {Reservoir} {Computing} on a {Photonic} {Chip} {Using}
  {Silicon}-{Based} {Ring} {Resonators}},}\ }\href {\doibase
  10.1109/JSTQE.2018.2836985} {\bibfield  {journal} {\bibinfo  {journal} {IEEE
  Journal of Selected Topics in Quantum Electronics}\ }\textbf {\bibinfo
  {volume} {24}},\ \bibinfo {pages} {1--8} (\bibinfo {year}
  {2018})}\BibitemShut {NoStop}%
\bibitem [{\citenamefont {Cai}, \citenamefont {Vasudev},\ and\ \citenamefont
  {Brongersma}(2011)}]{cai_electrically_2011}%
  \BibitemOpen
  \bibfield  {author} {\bibinfo {author} {\bibfnamefont {W.}~\bibnamefont
  {Cai}}, \bibinfo {author} {\bibfnamefont {A.~P.}\ \bibnamefont {Vasudev}}, \
  and\ \bibinfo {author} {\bibfnamefont {M.~L.}\ \bibnamefont {Brongersma}},\
  }\bibfield  {title} {\enquote {\bibinfo {title} {Electrically controlled
  nonlinear generation of light with plasmonics},}\ }\href {\doibase
  10.1126/science.1207858} {\bibfield  {journal} {\bibinfo  {journal} {Science
  (New York, N.Y.)}\ }\textbf {\bibinfo {volume} {333}},\ \bibinfo {pages}
  {1720--1723} (\bibinfo {year} {2011})}\BibitemShut {NoStop}%
\bibitem [{\citenamefont {Kauranen}\ and\ \citenamefont
  {Zayats}(2012)}]{kauranen_nonlinear_2012}%
  \BibitemOpen
  \bibfield  {author} {\bibinfo {author} {\bibfnamefont {M.}~\bibnamefont
  {Kauranen}}\ and\ \bibinfo {author} {\bibfnamefont {A.~V.}\ \bibnamefont
  {Zayats}},\ }\bibfield  {title} {\enquote {\bibinfo {title} {Nonlinear
  plasmonics},}\ }\href {\doibase 10.1038/nphoton.2012.244} {\bibfield
  {journal} {\bibinfo  {journal} {Nature Photonics}\ }\textbf {\bibinfo
  {volume} {6}},\ \bibinfo {pages} {737--748} (\bibinfo {year}
  {2012})}\BibitemShut {NoStop}%
\bibitem [{\citenamefont {Tait}\ \emph
  {et~al.}(2017{\natexlab{a}})\citenamefont {Tait}, \citenamefont {de~Lima},
  \citenamefont {Zhou}, \citenamefont {Wu}, \citenamefont {Nahmias},
  \citenamefont {Shastri},\ and\ \citenamefont
  {Prucnal}}]{tait_neuromorphic_2017}%
  \BibitemOpen
  \bibfield  {author} {\bibinfo {author} {\bibfnamefont {A.~N.}\ \bibnamefont
  {Tait}}, \bibinfo {author} {\bibfnamefont {T.~F.}\ \bibnamefont {de~Lima}},
  \bibinfo {author} {\bibfnamefont {E.}~\bibnamefont {Zhou}}, \bibinfo {author}
  {\bibfnamefont {A.~X.}\ \bibnamefont {Wu}}, \bibinfo {author} {\bibfnamefont
  {M.~A.}\ \bibnamefont {Nahmias}}, \bibinfo {author} {\bibfnamefont {B.~J.}\
  \bibnamefont {Shastri}}, \ and\ \bibinfo {author} {\bibfnamefont {P.~R.}\
  \bibnamefont {Prucnal}},\ }\bibfield  {title} {\enquote {\bibinfo {title}
  {Neuromorphic silicon photonic networks},}\ }\href {\doibase
  10.1038/s41598-017-07754-z} {\bibfield  {journal} {\bibinfo  {journal}
  {Scientific Reports}\ }\textbf {\bibinfo {volume} {7}} (\bibinfo {year}
  {2017}{\natexlab{a}}),\ 10.1038/s41598-017-07754-z},\ \bibinfo {note} {arXiv:
  1611.02272}\BibitemShut {NoStop}%
\bibitem [{\citenamefont {George}\ \emph {et~al.}(2018)\citenamefont {George},
  \citenamefont {Amin}, \citenamefont {Mehrabian}, \citenamefont {Khurgin},
  \citenamefont {El-Ghazawi}, \citenamefont {Prucnal},\ and\ \citenamefont
  {Sorger}}]{george_electrooptic_2018}%
  \BibitemOpen
  \bibfield  {author} {\bibinfo {author} {\bibfnamefont {J.}~\bibnamefont
  {George}}, \bibinfo {author} {\bibfnamefont {R.}~\bibnamefont {Amin}},
  \bibinfo {author} {\bibfnamefont {A.}~\bibnamefont {Mehrabian}}, \bibinfo
  {author} {\bibfnamefont {J.}~\bibnamefont {Khurgin}}, \bibinfo {author}
  {\bibfnamefont {T.}~\bibnamefont {El-Ghazawi}}, \bibinfo {author}
  {\bibfnamefont {P.~R.}\ \bibnamefont {Prucnal}}, \ and\ \bibinfo {author}
  {\bibfnamefont {V.~J.}\ \bibnamefont {Sorger}},\ }\bibfield  {title}
  {\enquote {\bibinfo {title} {Electrooptic {Nonlinear} {Activation}
  {Functions} for {Vector} {Matrix} {Multiplications} in {Optical} {Neural}
  {Networks}},}\ }in\ \href {\doibase 10.1364/SPPCOM.2018.SpW4G.3} {\emph
  {\bibinfo {booktitle} {Advanced {Photonics} 2018 ({BGPP}, {IPR}, {NP},
  {NOMA}, {Sensors}, {Networks}, {SPPCom}, {SOF}) (2018), paper {SpW}4G.3}}}\
  (\bibinfo  {publisher} {Optical Society of America},\ \bibinfo {year}
  {2018})\ p.\ \bibinfo {pages} {SpW4G.3}\BibitemShut {NoStop}%
\bibitem [{\citenamefont {Tait}\ \emph
  {et~al.}(2017{\natexlab{b}})\citenamefont {Tait}, \citenamefont {Lima},
  \citenamefont {Zhou}, \citenamefont {Wu}, \citenamefont {Nahmias},
  \citenamefont {Shastri},\ and\ \citenamefont
  {Prucnal}}]{tait_neuromorphic_2017-1}%
  \BibitemOpen
  \bibfield  {author} {\bibinfo {author} {\bibfnamefont {A.~N.}\ \bibnamefont
  {Tait}}, \bibinfo {author} {\bibfnamefont {T.~F.~d.}\ \bibnamefont {Lima}},
  \bibinfo {author} {\bibfnamefont {E.}~\bibnamefont {Zhou}}, \bibinfo {author}
  {\bibfnamefont {A.~X.}\ \bibnamefont {Wu}}, \bibinfo {author} {\bibfnamefont
  {M.~A.}\ \bibnamefont {Nahmias}}, \bibinfo {author} {\bibfnamefont {B.~J.}\
  \bibnamefont {Shastri}}, \ and\ \bibinfo {author} {\bibfnamefont {P.~R.}\
  \bibnamefont {Prucnal}},\ }\bibfield  {title} {\enquote {\bibinfo {title}
  {Neuromorphic photonic networks using silicon photonic weight banks},}\
  }\href {\doibase 10.1038/s41598-017-07754-z} {\bibfield  {journal} {\bibinfo
  {journal} {Scientific Reports}\ }\textbf {\bibinfo {volume} {7}},\ \bibinfo
  {pages} {7430} (\bibinfo {year} {2017}{\natexlab{b}})}\BibitemShut {NoStop}%
\bibitem [{\citenamefont {Yoo}\ \emph {et~al.}(2016)\citenamefont {Yoo},
  \citenamefont {Kim}, \citenamefont {Park},\ and\ \citenamefont
  {Nam}}]{yoo_electro-optical_2016}%
  \BibitemOpen
  \bibfield  {author} {\bibinfo {author} {\bibfnamefont {S.}~\bibnamefont
  {Yoo}}, \bibinfo {author} {\bibfnamefont {R.}~\bibnamefont {Kim}}, \bibinfo
  {author} {\bibfnamefont {J.-H.}\ \bibnamefont {Park}}, \ and\ \bibinfo
  {author} {\bibfnamefont {Y.}~\bibnamefont {Nam}},\ }\bibfield  {title}
  {\enquote {\bibinfo {title} {Electro-optical {Neural} {Platform} {Integrated}
  with {Nanoplasmonic} {Inhibition} {Interface}},}\ }\href {\doibase
  10.1021/acsnano.5b07747} {\bibfield  {journal} {\bibinfo  {journal} {ACS
  Nano}\ }\textbf {\bibinfo {volume} {10}},\ \bibinfo {pages} {4274--4281}
  (\bibinfo {year} {2016})}\BibitemShut {NoStop}%
\bibitem [{\citenamefont {Amin}, \citenamefont {Khurgin},\ and\ \citenamefont
  {Sorger}(2018)}]{amin_waveguide-based_2018}%
  \BibitemOpen
  \bibfield  {author} {\bibinfo {author} {\bibfnamefont {R.}~\bibnamefont
  {Amin}}, \bibinfo {author} {\bibfnamefont {J.~B.}\ \bibnamefont {Khurgin}}, \
  and\ \bibinfo {author} {\bibfnamefont {V.~J.}\ \bibnamefont {Sorger}},\
  }\bibfield  {title} {\enquote {\bibinfo {title} {Waveguide-based
  electro-absorption modulator performance: comparative analysis},}\ }\href
  {\doibase 10.1364/OE.26.015445} {\bibfield  {journal} {\bibinfo  {journal}
  {Optics Express}\ }\textbf {\bibinfo {volume} {26}},\ \bibinfo {pages}
  {15445--15470} (\bibinfo {year} {2018})}\BibitemShut {NoStop}%
\bibitem [{\citenamefont {Oulton}\ \emph {et~al.}(2008)\citenamefont {Oulton},
  \citenamefont {Sorger}, \citenamefont {Genov}, \citenamefont {Pile},\ and\
  \citenamefont {Zhang}}]{oulton_hybrid_2008}%
  \BibitemOpen
  \bibfield  {author} {\bibinfo {author} {\bibfnamefont {R.~F.}\ \bibnamefont
  {Oulton}}, \bibinfo {author} {\bibfnamefont {V.~J.}\ \bibnamefont {Sorger}},
  \bibinfo {author} {\bibfnamefont {D.~A.}\ \bibnamefont {Genov}}, \bibinfo
  {author} {\bibfnamefont {D.~F.~P.}\ \bibnamefont {Pile}}, \ and\ \bibinfo
  {author} {\bibfnamefont {X.}~\bibnamefont {Zhang}},\ }\bibfield  {title}
  {\enquote {\bibinfo {title} {A hybrid plasmonic waveguide for subwavelength
  confinement and long-range propagation},}\ }\href {\doibase
  10.1038/nphoton.2008.131} {\bibfield  {journal} {\bibinfo  {journal} {Nature
  Photonics}\ }\textbf {\bibinfo {volume} {2}},\ \bibinfo {pages} {496--500}
  (\bibinfo {year} {2008})}\BibitemShut {NoStop}%
\bibitem [{\citenamefont {Sorger}\ \emph {et~al.}(2012)\citenamefont {Sorger},
  \citenamefont {Lanzillotti-Kimura}, \citenamefont {Ma},\ and\ \citenamefont
  {Zhang}}]{sorger_ultra-compact_2012}%
  \BibitemOpen
  \bibfield  {author} {\bibinfo {author} {\bibfnamefont {V.~J.}\ \bibnamefont
  {Sorger}}, \bibinfo {author} {\bibfnamefont {N.~D.}\ \bibnamefont
  {Lanzillotti-Kimura}}, \bibinfo {author} {\bibfnamefont {R.-M.}\ \bibnamefont
  {Ma}}, \ and\ \bibinfo {author} {\bibfnamefont {X.}~\bibnamefont {Zhang}},\
  }\bibfield  {title} {\enquote {\bibinfo {title} {Ultra-compact silicon
  nanophotonic modulator with broadband response},}\ }\href {\doibase
  10.1515/nanoph-2012-0009} {\bibfield  {journal} {\bibinfo  {journal}
  {Nanophotonics}\ }\textbf {\bibinfo {volume} {1}},\ \bibinfo {pages} {17--22}
  (\bibinfo {year} {2012})}\BibitemShut {NoStop}%
\bibitem [{\citenamefont {Haffner}\ \emph {et~al.}(2018)\citenamefont
  {Haffner}, \citenamefont {Chelladurai}, \citenamefont {Fedoryshyn},
  \citenamefont {Josten}, \citenamefont {Baeuerle}, \citenamefont {Heni},
  \citenamefont {Watanabe}, \citenamefont {Cui}, \citenamefont {Cheng},
  \citenamefont {Saha}, \citenamefont {Elder}, \citenamefont {Dalton},
  \citenamefont {Boltasseva}, \citenamefont {Shalaev}, \citenamefont {Kinsey},\
  and\ \citenamefont {Leuthold}}]{haffner_low-loss_2018}%
  \BibitemOpen
  \bibfield  {author} {\bibinfo {author} {\bibfnamefont {C.}~\bibnamefont
  {Haffner}}, \bibinfo {author} {\bibfnamefont {D.}~\bibnamefont
  {Chelladurai}}, \bibinfo {author} {\bibfnamefont {Y.}~\bibnamefont
  {Fedoryshyn}}, \bibinfo {author} {\bibfnamefont {A.}~\bibnamefont {Josten}},
  \bibinfo {author} {\bibfnamefont {B.}~\bibnamefont {Baeuerle}}, \bibinfo
  {author} {\bibfnamefont {W.}~\bibnamefont {Heni}}, \bibinfo {author}
  {\bibfnamefont {T.}~\bibnamefont {Watanabe}}, \bibinfo {author}
  {\bibfnamefont {T.}~\bibnamefont {Cui}}, \bibinfo {author} {\bibfnamefont
  {B.}~\bibnamefont {Cheng}}, \bibinfo {author} {\bibfnamefont
  {S.}~\bibnamefont {Saha}}, \bibinfo {author} {\bibfnamefont {D.~L.}\
  \bibnamefont {Elder}}, \bibinfo {author} {\bibfnamefont {L.~R.}\ \bibnamefont
  {Dalton}}, \bibinfo {author} {\bibfnamefont {A.}~\bibnamefont {Boltasseva}},
  \bibinfo {author} {\bibfnamefont {V.~M.}\ \bibnamefont {Shalaev}}, \bibinfo
  {author} {\bibfnamefont {N.}~\bibnamefont {Kinsey}}, \ and\ \bibinfo {author}
  {\bibfnamefont {J.}~\bibnamefont {Leuthold}},\ }\bibfield  {title} {\enquote
  {\bibinfo {title} {Low-loss plasmon-assisted electro-optic modulator},}\
  }\href {\doibase 10.1038/s41586-018-0031-4} {\bibfield  {journal} {\bibinfo
  {journal} {Nature}\ }\textbf {\bibinfo {volume} {556}},\ \bibinfo {pages}
  {483} (\bibinfo {year} {2018})}\BibitemShut {NoStop}%
\bibitem [{\citenamefont {Keeler}\ \emph {et~al.}(2017)\citenamefont {Keeler},
  \citenamefont {Geib}, \citenamefont {Serkland}, \citenamefont {Parameswaran},
  \citenamefont {Luk}, \citenamefont {Luk}, \citenamefont {Griñe},
  \citenamefont {Ihlefeld}, \citenamefont {Campione},\ and\ \citenamefont
  {Wendt}}]{keeler_multi-gigabit_2017}%
  \BibitemOpen
  \bibfield  {author} {\bibinfo {author} {\bibfnamefont {G.~A.}\ \bibnamefont
  {Keeler}}, \bibinfo {author} {\bibfnamefont {K.~M.}\ \bibnamefont {Geib}},
  \bibinfo {author} {\bibfnamefont {D.~K.}\ \bibnamefont {Serkland}}, \bibinfo
  {author} {\bibfnamefont {S.}~\bibnamefont {Parameswaran}}, \bibinfo {author}
  {\bibfnamefont {T.~S.}\ \bibnamefont {Luk}}, \bibinfo {author} {\bibfnamefont
  {T.~S.}\ \bibnamefont {Luk}}, \bibinfo {author} {\bibfnamefont {A.~J.}\
  \bibnamefont {Griñe}}, \bibinfo {author} {\bibfnamefont {J.}~\bibnamefont
  {Ihlefeld}}, \bibinfo {author} {\bibfnamefont {S.}~\bibnamefont {Campione}},
  \ and\ \bibinfo {author} {\bibfnamefont {J.~R.}\ \bibnamefont {Wendt}},\
  }\bibfield  {title} {\enquote {\bibinfo {title} {Multi-{Gigabit} {Operation}
  of a {Compact}, {Broadband} {Modulator} {Based} on {ENZ} {Confinement} in
  {Indium} {Oxide}},}\ }in\ \href {\doibase 10.1364/OFC.2017.Th3I.1} {\emph
  {\bibinfo {booktitle} {Optical {Fiber} {Communication} {Conference} (2017),
  paper {Th}3I.1}}}\ (\bibinfo  {publisher} {Optical Society of America},\
  \bibinfo {year} {2017})\ p.\ \bibinfo {pages} {Th3I.1}\BibitemShut {NoStop}%
\bibitem [{\citenamefont {Zhu}, \citenamefont {Lo},\ and\ \citenamefont
  {Kwong}(2013)}]{zhu_phase_2013}%
  \BibitemOpen
  \bibfield  {author} {\bibinfo {author} {\bibfnamefont {S.}~\bibnamefont
  {Zhu}}, \bibinfo {author} {\bibfnamefont {G.~Q.}\ \bibnamefont {Lo}}, \ and\
  \bibinfo {author} {\bibfnamefont {D.~L.}\ \bibnamefont {Kwong}},\ }\bibfield
  {title} {\enquote {\bibinfo {title} {Phase modulation in horizontal
  metal-insulator-silicon-insulator-metal plasmonic waveguides},}\ }\href
  {\doibase 10.1364/OE.21.008320} {\bibfield  {journal} {\bibinfo  {journal}
  {Optics Express}\ }\textbf {\bibinfo {volume} {21}},\ \bibinfo {pages}
  {8320--8330} (\bibinfo {year} {2013})}\BibitemShut {NoStop}%
\bibitem [{\citenamefont {Wu}\ \emph {et~al.}(2008)\citenamefont {Wu},
  \citenamefont {Nelson}, \citenamefont {Haus},\ and\ \citenamefont
  {Zhan}}]{wu_plasmonic_2008}%
  \BibitemOpen
  \bibfield  {author} {\bibinfo {author} {\bibfnamefont {Z.}~\bibnamefont
  {Wu}}, \bibinfo {author} {\bibfnamefont {R.~L.}\ \bibnamefont {Nelson}},
  \bibinfo {author} {\bibfnamefont {J.~W.}\ \bibnamefont {Haus}}, \ and\
  \bibinfo {author} {\bibfnamefont {Q.}~\bibnamefont {Zhan}},\ }\bibfield
  {title} {\enquote {\bibinfo {title} {Plasmonic electro-optic modulator design
  using a resonant metal grating},}\ }\href {\doibase 10.1364/OL.33.000551}
  {\bibfield  {journal} {\bibinfo  {journal} {Optics Letters}\ }\textbf
  {\bibinfo {volume} {33}},\ \bibinfo {pages} {551--553} (\bibinfo {year}
  {2008})}\BibitemShut {NoStop}%
\bibitem [{\citenamefont {Gan}\ \emph {et~al.}(2013)\citenamefont {Gan},
  \citenamefont {Shiue}, \citenamefont {Gao}, \citenamefont {Mak},
  \citenamefont {Yao}, \citenamefont {Li}, \citenamefont {Szep}, \citenamefont
  {Walker}, \citenamefont {Hone}, \citenamefont {Heinz},\ and\ \citenamefont
  {Englund}}]{gan_high-contrast_2013}%
  \BibitemOpen
  \bibfield  {author} {\bibinfo {author} {\bibfnamefont {X.}~\bibnamefont
  {Gan}}, \bibinfo {author} {\bibfnamefont {R.-J.}\ \bibnamefont {Shiue}},
  \bibinfo {author} {\bibfnamefont {Y.}~\bibnamefont {Gao}}, \bibinfo {author}
  {\bibfnamefont {K.~F.}\ \bibnamefont {Mak}}, \bibinfo {author} {\bibfnamefont
  {X.}~\bibnamefont {Yao}}, \bibinfo {author} {\bibfnamefont {L.}~\bibnamefont
  {Li}}, \bibinfo {author} {\bibfnamefont {A.}~\bibnamefont {Szep}}, \bibinfo
  {author} {\bibfnamefont {D.}~\bibnamefont {Walker}}, \bibinfo {author}
  {\bibfnamefont {J.}~\bibnamefont {Hone}}, \bibinfo {author} {\bibfnamefont
  {T.~F.}\ \bibnamefont {Heinz}}, \ and\ \bibinfo {author} {\bibfnamefont
  {D.}~\bibnamefont {Englund}},\ }\bibfield  {title} {\enquote {\bibinfo
  {title} {High-{Contrast} {Electrooptic} {Modulation} of a {Photonic}
  {Crystal} {Nanocavity} by {Electrical} {Gating} of {Graphene}},}\ }\href
  {\doibase 10.1021/nl304357u} {\bibfield  {journal} {\bibinfo  {journal} {Nano
  Letters}\ }\textbf {\bibinfo {volume} {13}},\ \bibinfo {pages} {691--696}
  (\bibinfo {year} {2013})}\BibitemShut {NoStop}%
\bibitem [{\citenamefont {Briggs}\ \emph {et~al.}(2010)\citenamefont {Briggs},
  \citenamefont {Grandidier}, \citenamefont {Burgos}, \citenamefont
  {Feigenbaum},\ and\ \citenamefont {Atwater}}]{briggs_efficient_2010}%
  \BibitemOpen
  \bibfield  {author} {\bibinfo {author} {\bibfnamefont {R.~M.}\ \bibnamefont
  {Briggs}}, \bibinfo {author} {\bibfnamefont {J.}~\bibnamefont {Grandidier}},
  \bibinfo {author} {\bibfnamefont {S.~P.}\ \bibnamefont {Burgos}}, \bibinfo
  {author} {\bibfnamefont {E.}~\bibnamefont {Feigenbaum}}, \ and\ \bibinfo
  {author} {\bibfnamefont {H.~A.}\ \bibnamefont {Atwater}},\ }\bibfield
  {title} {\enquote {\bibinfo {title} {Efficient {Coupling} between
  {Dielectric}-{Loaded} {Plasmonic} and {Silicon} {Photonic} {Waveguides}},}\
  }\href {\doibase 10.1021/nl1024529} {\bibfield  {journal} {\bibinfo
  {journal} {Nano Letters}\ }\textbf {\bibinfo {volume} {10}},\ \bibinfo
  {pages} {4851--4857} (\bibinfo {year} {2010})}\BibitemShut {NoStop}%
\bibitem [{\citenamefont {Lee}\ \emph {et~al.}(2014)\citenamefont {Lee},
  \citenamefont {Papadakis}, \citenamefont {Burgos}, \citenamefont {Chander},
  \citenamefont {Kriesch}, \citenamefont {Pala}, \citenamefont {Peschel},\ and\
  \citenamefont {Atwater}}]{lee_nanoscale_2014}%
  \BibitemOpen
  \bibfield  {author} {\bibinfo {author} {\bibfnamefont {H.~W.}\ \bibnamefont
  {Lee}}, \bibinfo {author} {\bibfnamefont {G.}~\bibnamefont {Papadakis}},
  \bibinfo {author} {\bibfnamefont {S.~P.}\ \bibnamefont {Burgos}}, \bibinfo
  {author} {\bibfnamefont {K.}~\bibnamefont {Chander}}, \bibinfo {author}
  {\bibfnamefont {A.}~\bibnamefont {Kriesch}}, \bibinfo {author} {\bibfnamefont
  {R.}~\bibnamefont {Pala}}, \bibinfo {author} {\bibfnamefont {U.}~\bibnamefont
  {Peschel}}, \ and\ \bibinfo {author} {\bibfnamefont {H.~A.}\ \bibnamefont
  {Atwater}},\ }\bibfield  {title} {\enquote {\bibinfo {title} {Nanoscale
  {Conducting} {Oxide} {PlasMOStor}},}\ }\href {\doibase 10.1021/nl502998z}
  {\bibfield  {journal} {\bibinfo  {journal} {Nano Letters}\ }\textbf {\bibinfo
  {volume} {14}},\ \bibinfo {pages} {6463--6468} (\bibinfo {year}
  {2014})}\BibitemShut {NoStop}%
\bibitem [{\citenamefont {Vasudev}\ \emph {et~al.}(2013)\citenamefont
  {Vasudev}, \citenamefont {Kang}, \citenamefont {Park}, \citenamefont {Liu},\
  and\ \citenamefont {Brongersma}}]{vasudev_electro-optical_2013}%
  \BibitemOpen
  \bibfield  {author} {\bibinfo {author} {\bibfnamefont {A.~P.}\ \bibnamefont
  {Vasudev}}, \bibinfo {author} {\bibfnamefont {J.-H.}\ \bibnamefont {Kang}},
  \bibinfo {author} {\bibfnamefont {J.}~\bibnamefont {Park}}, \bibinfo {author}
  {\bibfnamefont {X.}~\bibnamefont {Liu}}, \ and\ \bibinfo {author}
  {\bibfnamefont {M.~L.}\ \bibnamefont {Brongersma}},\ }\bibfield  {title}
  {\enquote {\bibinfo {title} {Electro-optical modulation of a silicon
  waveguide with an “epsilon-near-zero” material},}\ }\href {\doibase
  10.1364/OE.21.026387} {\bibfield  {journal} {\bibinfo  {journal} {Optics
  Express}\ }\textbf {\bibinfo {volume} {21}},\ \bibinfo {pages} {26387--26397}
  (\bibinfo {year} {2013})}\BibitemShut {NoStop}%
\bibitem [{\citenamefont {Babicheva}, \citenamefont {Boltasseva},\ and\
  \citenamefont {Lavrinenko}(2015)}]{babicheva_transparent_2015}%
  \BibitemOpen
  \bibfield  {author} {\bibinfo {author} {\bibfnamefont {V.~E.}\ \bibnamefont
  {Babicheva}}, \bibinfo {author} {\bibfnamefont {A.}~\bibnamefont
  {Boltasseva}}, \ and\ \bibinfo {author} {\bibfnamefont {A.~V.}\ \bibnamefont
  {Lavrinenko}},\ }\bibfield  {title} {\enquote {\bibinfo {title} {Transparent
  conducting oxides for electro-optical plasmonic modulators},}\ }\href
  {\doibase 10.1515/nanoph-2015-0004} {\bibfield  {journal} {\bibinfo
  {journal} {Nanophotonics}\ }\textbf {\bibinfo {volume} {4}},\ \bibinfo
  {pages} {165--185} (\bibinfo {year} {2015})}\BibitemShut {NoStop}%
\bibitem [{\citenamefont {Abdelatty}, \citenamefont {Badr},\ and\ \citenamefont
  {Swillam}(2018)}]{abdelatty_hybrid_2018}%
  \BibitemOpen
  \bibfield  {author} {\bibinfo {author} {\bibfnamefont {M.~Y.}\ \bibnamefont
  {Abdelatty}}, \bibinfo {author} {\bibfnamefont {M.~M.}\ \bibnamefont {Badr}},
  \ and\ \bibinfo {author} {\bibfnamefont {M.~A.}\ \bibnamefont {Swillam}},\
  }\bibfield  {title} {\enquote {\bibinfo {title} {Hybrid plasmonic
  electro-optical absorption modulator based on epsilon-near-zero
  characteristics of {ITO}},}\ }in\ \href {\doibase 10.1117/12.2288912} {\emph
  {\bibinfo {booktitle} {Integrated {Optics}: {Devices}, {Materials}, and
  {Technologies} {XXII}}}},\ Vol.\ \bibinfo {volume} {10535}\ (\bibinfo
  {publisher} {International Society for Optics and Photonics},\ \bibinfo
  {year} {2018})\ p.\ \bibinfo {pages} {105351T}\BibitemShut {NoStop}%
\bibitem [{\citenamefont {Alam}, \citenamefont {Leon},\ and\ \citenamefont
  {Boyd}(2016)}]{alam_large_2016}%
  \BibitemOpen
  \bibfield  {author} {\bibinfo {author} {\bibfnamefont {M.~Z.}\ \bibnamefont
  {Alam}}, \bibinfo {author} {\bibfnamefont {I.~D.}\ \bibnamefont {Leon}}, \
  and\ \bibinfo {author} {\bibfnamefont {R.~W.}\ \bibnamefont {Boyd}},\
  }\bibfield  {title} {\enquote {\bibinfo {title} {Large optical nonlinearity
  of indium tin oxide in its epsilon-near-zero region},}\ }\href {\doibase
  10.1126/science.aae0330} {\bibfield  {journal} {\bibinfo  {journal}
  {Science}\ }\textbf {\bibinfo {volume} {352}},\ \bibinfo {pages} {795--797}
  (\bibinfo {year} {2016})}\BibitemShut {NoStop}%
\bibitem [{\citenamefont {Zhao}\ \emph {et~al.}(2014)\citenamefont {Zhao},
  \citenamefont {Wang}, \citenamefont {Capretti}, \citenamefont {Negro},\ and\
  \citenamefont {Klamkin}}]{zhao_silicon_2014}%
  \BibitemOpen
  \bibfield  {author} {\bibinfo {author} {\bibfnamefont {H.}~\bibnamefont
  {Zhao}}, \bibinfo {author} {\bibfnamefont {Y.}~\bibnamefont {Wang}}, \bibinfo
  {author} {\bibfnamefont {A.}~\bibnamefont {Capretti}}, \bibinfo {author}
  {\bibfnamefont {L.~D.}\ \bibnamefont {Negro}}, \ and\ \bibinfo {author}
  {\bibfnamefont {J.}~\bibnamefont {Klamkin}},\ }\bibfield  {title} {\enquote
  {\bibinfo {title} {Silicon photonic modulators based on epsilon-near-zero
  indium tin oxide materials},}\ }in\ \href {\doibase
  10.1109/IPCon.2014.6995380} {\emph {\bibinfo {booktitle} {2014 {IEEE}
  {Photonics} {Conference}}}}\ (\bibinfo {year} {2014})\ pp.\ \bibinfo {pages}
  {334--335}\BibitemShut {NoStop}%
\bibitem [{noa(2008)}]{noauthor_us8049862b2_nodate}%
  \BibitemOpen
  \href {https://patents.google.com/patent/US8049862B2/en} {\enquote {\bibinfo
  {title} {{US}8049862b2 - {Indium} tin oxide ({ITO}) layer forming - {Google}
  {Patents}},}\ } (\bibinfo {year} {2008})\BibitemShut {NoStop}%
\bibitem [{\citenamefont {Kerkache}\ \emph {et~al.}(2006)\citenamefont
  {Kerkache}, \citenamefont {Layadi}, \citenamefont {Dogheche},\ and\
  \citenamefont {Rémiens}}]{kerkache_physical_2006}%
  \BibitemOpen
  \bibfield  {author} {\bibinfo {author} {\bibfnamefont {L.}~\bibnamefont
  {Kerkache}}, \bibinfo {author} {\bibfnamefont {A.}~\bibnamefont {Layadi}},
  \bibinfo {author} {\bibfnamefont {E.}~\bibnamefont {Dogheche}}, \ and\
  \bibinfo {author} {\bibfnamefont {D.}~\bibnamefont {Rémiens}},\ }\bibfield
  {title} {\enquote {\bibinfo {title} {Physical properties of {RF} sputtered
  {ITO} thin films and annealing effect},}\ }\href {\doibase
  10.1088/0022-3727/39/1/027} {\bibfield  {journal} {\bibinfo  {journal}
  {Journal of Physics D: Applied Physics}\ }\textbf {\bibinfo {volume} {39}},\
  \bibinfo {pages} {184} (\bibinfo {year} {2006})}\BibitemShut {NoStop}%
\bibitem [{\citenamefont {Chityuttakan}\ \emph {et~al.}(2019)\citenamefont
  {Chityuttakan}, \citenamefont {Chinvetkitvanich}, \citenamefont
  {Chatraphorn},\ and\ \citenamefont
  {Chatraphorn}}]{chityuttakan_influence_nodate}%
  \BibitemOpen
  \bibfield  {author} {\bibinfo {author} {\bibfnamefont {C.}~\bibnamefont
  {Chityuttakan}}, \bibinfo {author} {\bibfnamefont {P.}~\bibnamefont
  {Chinvetkitvanich}}, \bibinfo {author} {\bibfnamefont {S.}~\bibnamefont
  {Chatraphorn}}, \ and\ \bibinfo {author} {\bibfnamefont {S.}~\bibnamefont
  {Chatraphorn}},\ }\bibfield  {title} {\enquote {\bibinfo {title} {Influence
  of deposition parameters on the quality of {ITO} films for photovoltaic
  application},}\ }\href@noop {} {\bibfield  {journal} {\bibinfo  {journal}
  {AIP Conference Proceedings 2091}\ ,\ \bibinfo {pages} {3}} (\bibinfo {year}
  {2019})}\BibitemShut {NoStop}%
\bibitem [{\citenamefont {Hu}\ \emph {et~al.}(2004)\citenamefont {Hu},
  \citenamefont {Diao}, \citenamefont {Wang}, \citenamefont {Hao},\ and\
  \citenamefont {Wang}}]{hu_effects_2004}%
  \BibitemOpen
  \bibfield  {author} {\bibinfo {author} {\bibfnamefont {Y.}~\bibnamefont
  {Hu}}, \bibinfo {author} {\bibfnamefont {X.}~\bibnamefont {Diao}}, \bibinfo
  {author} {\bibfnamefont {C.}~\bibnamefont {Wang}}, \bibinfo {author}
  {\bibfnamefont {W.}~\bibnamefont {Hao}}, \ and\ \bibinfo {author}
  {\bibfnamefont {T.}~\bibnamefont {Wang}},\ }\bibfield  {title} {\enquote
  {\bibinfo {title} {Effects of heat treatment on properties of {ITO} films
  prepared by rf magnetron sputtering},}\ }\href {\doibase
  10.1016/j.vacuum.2004.01.081} {\bibfield  {journal} {\bibinfo  {journal}
  {Vacuum}\ }\textbf {\bibinfo {volume} {75}},\ \bibinfo {pages} {183--188}
  (\bibinfo {year} {2004})}\BibitemShut {NoStop}%
\bibitem [{\citenamefont {Bhatti}, \citenamefont {Rana},\ and\ \citenamefont
  {Khan}(2004)}]{bhatti_characterization_2004}%
  \BibitemOpen
  \bibfield  {author} {\bibinfo {author} {\bibfnamefont {M.~T.}\ \bibnamefont
  {Bhatti}}, \bibinfo {author} {\bibfnamefont {A.~M.}\ \bibnamefont {Rana}}, \
  and\ \bibinfo {author} {\bibfnamefont {A.~F.}\ \bibnamefont {Khan}},\
  }\bibfield  {title} {\enquote {\bibinfo {title} {Characterization of
  rf-sputtered indium tin oxide thin films},}\ }\href {\doibase
  10.1016/j.matchemphys.2003.11.022} {\bibfield  {journal} {\bibinfo  {journal}
  {Materials Chemistry and Physics}\ }\textbf {\bibinfo {volume} {84}},\
  \bibinfo {pages} {126--130} (\bibinfo {year} {2004})}\BibitemShut {NoStop}%
\bibitem [{\citenamefont {Lee}\ and\ \citenamefont
  {Ok~Park}(2004)}]{lee_behaviors_2004}%
  \BibitemOpen
  \bibfield  {author} {\bibinfo {author} {\bibfnamefont {H.-C.}\ \bibnamefont
  {Lee}}\ and\ \bibinfo {author} {\bibfnamefont {O.}~\bibnamefont {Ok~Park}},\
  }\bibfield  {title} {\enquote {\bibinfo {title} {Behaviors of carrier
  concentrations and mobilities in indium–tin oxide thin films by {DC}
  magnetron sputtering at various oxygen flow rates},}\ }\href {\doibase
  10.1016/j.vacuum.2004.08.006} {\bibfield  {journal} {\bibinfo  {journal}
  {Vacuum}\ }\textbf {\bibinfo {volume} {77}},\ \bibinfo {pages} {69--77}
  (\bibinfo {year} {2004})}\BibitemShut {NoStop}%
\bibitem [{\citenamefont {Amin}\ \emph
  {et~al.}(2018{\natexlab{b}})\citenamefont {Amin}, \citenamefont {Maiti},
  \citenamefont {Carfano}, \citenamefont {Ma}, \citenamefont {Tahersima},
  \citenamefont {Lilach}, \citenamefont {Ratnayake}, \citenamefont {Dalir},\
  and\ \citenamefont {Sorger}}]{amin_0.52_2018}%
  \BibitemOpen
  \bibfield  {author} {\bibinfo {author} {\bibfnamefont {R.}~\bibnamefont
  {Amin}}, \bibinfo {author} {\bibfnamefont {R.}~\bibnamefont {Maiti}},
  \bibinfo {author} {\bibfnamefont {C.}~\bibnamefont {Carfano}}, \bibinfo
  {author} {\bibfnamefont {Z.}~\bibnamefont {Ma}}, \bibinfo {author}
  {\bibfnamefont {M.~H.}\ \bibnamefont {Tahersima}}, \bibinfo {author}
  {\bibfnamefont {Y.}~\bibnamefont {Lilach}}, \bibinfo {author} {\bibfnamefont
  {D.}~\bibnamefont {Ratnayake}}, \bibinfo {author} {\bibfnamefont
  {H.}~\bibnamefont {Dalir}}, \ and\ \bibinfo {author} {\bibfnamefont {V.~J.}\
  \bibnamefont {Sorger}},\ }\bibfield  {title} {\enquote {\bibinfo {title}
  {0.52 {V} mm {ITO}-based {Mach}-{Zehnder} modulator in silicon photonics},}\
  }\href {\doibase 10.1063/1.5052635} {\bibfield  {journal} {\bibinfo
  {journal} {APL Photonics}\ }\textbf {\bibinfo {volume} {3}},\ \bibinfo
  {pages} {126104} (\bibinfo {year} {2018}{\natexlab{b}})}\BibitemShut
  {NoStop}%
\bibitem [{\citenamefont {Gui}\ \emph {et~al.}(2018)\citenamefont {Gui},
  \citenamefont {Miscuglio}, \citenamefont {Ma}, \citenamefont {Tahersima},
  \citenamefont {Sun}, \citenamefont {Amin}, \citenamefont {Dalir},\ and\
  \citenamefont {Sorger}}]{gui_towards_2018}%
  \BibitemOpen
  \bibfield  {author} {\bibinfo {author} {\bibfnamefont {Y.}~\bibnamefont
  {Gui}}, \bibinfo {author} {\bibfnamefont {M.}~\bibnamefont {Miscuglio}},
  \bibinfo {author} {\bibfnamefont {Z.}~\bibnamefont {Ma}}, \bibinfo {author}
  {\bibfnamefont {M.~T.}\ \bibnamefont {Tahersima}}, \bibinfo {author}
  {\bibfnamefont {S.}~\bibnamefont {Sun}}, \bibinfo {author} {\bibfnamefont
  {R.}~\bibnamefont {Amin}}, \bibinfo {author} {\bibfnamefont {H.}~\bibnamefont
  {Dalir}}, \ and\ \bibinfo {author} {\bibfnamefont {V.~J.}\ \bibnamefont
  {Sorger}},\ }\bibfield  {title} {\enquote {\bibinfo {title} {Towards
  integrated metatronics: a holistic approach on precise optical and electrical
  properties of {Indium} {Tin} {Oxide}},}\ }\href
  {https://arxiv.org/abs/1811.08344v4} {\bibfield  {journal} {\bibinfo
  {journal} {arXiv}\ } (\bibinfo {year} {2018})}\BibitemShut {NoStop}%
\bibitem [{\citenamefont {Tang}, \citenamefont {Peters},\ and\ \citenamefont
  {Bowers}(2012)}]{tang_over_2012}%
  \BibitemOpen
  \bibfield  {author} {\bibinfo {author} {\bibfnamefont {Y.}~\bibnamefont
  {Tang}}, \bibinfo {author} {\bibfnamefont {J.~D.}\ \bibnamefont {Peters}}, \
  and\ \bibinfo {author} {\bibfnamefont {J.~E.}\ \bibnamefont {Bowers}},\
  }\bibfield  {title} {\enquote {\bibinfo {title} {Over 67 {GHz} bandwidth
  hybrid silicon electroabsorption modulator with asymmetric segmented
  electrode for 1.3 $\mu$m transmission},}\ }\href {\doibase
  10.1364/OE.20.011529} {\bibfield  {journal} {\bibinfo  {journal} {Optics
  Express}\ }\textbf {\bibinfo {volume} {20}},\ \bibinfo {pages} {11529--11535}
  (\bibinfo {year} {2012})}\BibitemShut {NoStop}%
\bibitem [{\citenamefont {Tu}\ \emph {et~al.}(2011)\citenamefont {Tu},
  \citenamefont {Liow}, \citenamefont {Song}, \citenamefont {Yu},\ and\
  \citenamefont {Lo}}]{tu_fabrication_2011}%
  \BibitemOpen
  \bibfield  {author} {\bibinfo {author} {\bibfnamefont {X.}~\bibnamefont
  {Tu}}, \bibinfo {author} {\bibfnamefont {T.-Y.}\ \bibnamefont {Liow}},
  \bibinfo {author} {\bibfnamefont {J.}~\bibnamefont {Song}}, \bibinfo {author}
  {\bibfnamefont {M.}~\bibnamefont {Yu}}, \ and\ \bibinfo {author}
  {\bibfnamefont {G.~Q.}\ \bibnamefont {Lo}},\ }\bibfield  {title} {\enquote
  {\bibinfo {title} {Fabrication of low loss and high speed silicon optical
  modulator using doping compensation method},}\ }\href {\doibase
  10.1364/OE.19.018029} {\bibfield  {journal} {\bibinfo  {journal} {Optics
  Express}\ }\textbf {\bibinfo {volume} {19}},\ \bibinfo {pages} {18029--18035}
  (\bibinfo {year} {2011})}\BibitemShut {NoStop}%
\bibitem [{\citenamefont {Soref}\ and\ \citenamefont
  {Bennett}(1987)}]{soref_electrooptical_1987}%
  \BibitemOpen
  \bibfield  {author} {\bibinfo {author} {\bibfnamefont {R.}~\bibnamefont
  {Soref}}\ and\ \bibinfo {author} {\bibfnamefont {B.}~\bibnamefont
  {Bennett}},\ }\bibfield  {title} {\enquote {\bibinfo {title} {Electrooptical
  effects in silicon},}\ }\href {\doibase 10.1109/JQE.1987.1073206} {\bibfield
  {journal} {\bibinfo  {journal} {IEEE Journal of Quantum Electronics}\
  }\textbf {\bibinfo {volume} {23}},\ \bibinfo {pages} {123--129} (\bibinfo
  {year} {1987})}\BibitemShut {NoStop}%
\bibitem [{\citenamefont {Reed}\ \emph {et~al.}(2013)\citenamefont {Reed},
  \citenamefont {Mashanovich}, \citenamefont {Gardes}, \citenamefont
  {Nedeljkovic}, \citenamefont {Hu}, \citenamefont {Thomson}, \citenamefont
  {Li}, \citenamefont {Wilson}, \citenamefont {Chen},\ and\ \citenamefont
  {Hsu}}]{reed_recent_2013}%
  \BibitemOpen
  \bibfield  {author} {\bibinfo {author} {\bibfnamefont {G.~T.}\ \bibnamefont
  {Reed}}, \bibinfo {author} {\bibfnamefont {G.~Z.}\ \bibnamefont
  {Mashanovich}}, \bibinfo {author} {\bibfnamefont {F.~Y.}\ \bibnamefont
  {Gardes}}, \bibinfo {author} {\bibfnamefont {M.}~\bibnamefont {Nedeljkovic}},
  \bibinfo {author} {\bibfnamefont {Y.}~\bibnamefont {Hu}}, \bibinfo {author}
  {\bibfnamefont {D.~J.}\ \bibnamefont {Thomson}}, \bibinfo {author}
  {\bibfnamefont {K.}~\bibnamefont {Li}}, \bibinfo {author} {\bibfnamefont
  {P.~R.}\ \bibnamefont {Wilson}}, \bibinfo {author} {\bibfnamefont {S.-W.}\
  \bibnamefont {Chen}}, \ and\ \bibinfo {author} {\bibfnamefont {S.~S.}\
  \bibnamefont {Hsu}},\ }\bibfield  {title} {\enquote {\bibinfo {title} {Recent
  breakthroughs in carrier depletion based silicon optical modulators},}\
  }\href {\doibase 10.1515/nanoph-2013-0016} {\bibfield  {journal} {\bibinfo
  {journal} {Nanophotonics}\ }\textbf {\bibinfo {volume} {3}},\ \bibinfo
  {pages} {229--245} (\bibinfo {year} {2013})}\BibitemShut {NoStop}%
\bibitem [{\citenamefont {Haffner}\ \emph {et~al.}(2015)\citenamefont
  {Haffner}, \citenamefont {Heni}, \citenamefont {Fedoryshyn}, \citenamefont
  {Niegemann}, \citenamefont {Melikyan}, \citenamefont {Elder}, \citenamefont
  {Baeuerle}, \citenamefont {Salamin}, \citenamefont {Josten}, \citenamefont
  {Koch}, \citenamefont {Hoessbacher}, \citenamefont {Ducry}, \citenamefont
  {Juchli}, \citenamefont {Emboras}, \citenamefont {Hillerkuss}, \citenamefont
  {Kohl}, \citenamefont {Dalton}, \citenamefont {Hafner},\ and\ \citenamefont
  {Leuthold}}]{haffner_all-plasmonic_2015}%
  \BibitemOpen
  \bibfield  {author} {\bibinfo {author} {\bibfnamefont {C.}~\bibnamefont
  {Haffner}}, \bibinfo {author} {\bibfnamefont {W.}~\bibnamefont {Heni}},
  \bibinfo {author} {\bibfnamefont {Y.}~\bibnamefont {Fedoryshyn}}, \bibinfo
  {author} {\bibfnamefont {J.}~\bibnamefont {Niegemann}}, \bibinfo {author}
  {\bibfnamefont {A.}~\bibnamefont {Melikyan}}, \bibinfo {author}
  {\bibfnamefont {D.~L.}\ \bibnamefont {Elder}}, \bibinfo {author}
  {\bibfnamefont {B.}~\bibnamefont {Baeuerle}}, \bibinfo {author}
  {\bibfnamefont {Y.}~\bibnamefont {Salamin}}, \bibinfo {author} {\bibfnamefont
  {A.}~\bibnamefont {Josten}}, \bibinfo {author} {\bibfnamefont
  {U.}~\bibnamefont {Koch}}, \bibinfo {author} {\bibfnamefont {C.}~\bibnamefont
  {Hoessbacher}}, \bibinfo {author} {\bibfnamefont {F.}~\bibnamefont {Ducry}},
  \bibinfo {author} {\bibfnamefont {L.}~\bibnamefont {Juchli}}, \bibinfo
  {author} {\bibfnamefont {A.}~\bibnamefont {Emboras}}, \bibinfo {author}
  {\bibfnamefont {D.}~\bibnamefont {Hillerkuss}}, \bibinfo {author}
  {\bibfnamefont {M.}~\bibnamefont {Kohl}}, \bibinfo {author} {\bibfnamefont
  {L.~R.}\ \bibnamefont {Dalton}}, \bibinfo {author} {\bibfnamefont
  {C.}~\bibnamefont {Hafner}}, \ and\ \bibinfo {author} {\bibfnamefont
  {J.}~\bibnamefont {Leuthold}},\ }\bibfield  {title} {\enquote {\bibinfo
  {title} {All-plasmonic {Mach}–{Zehnder} modulator enabling optical
  high-speed communication at the microscale},}\ }\href {\doibase
  10.1038/nphoton.2015.127} {\bibfield  {journal} {\bibinfo  {journal} {Nature
  Photonics}\ }\textbf {\bibinfo {volume} {9}},\ \bibinfo {pages} {525--528}
  (\bibinfo {year} {2015})}\BibitemShut {NoStop}%
\bibitem [{\citenamefont {Sze}\ and\ \citenamefont
  {Ng}(2006)}]{sze_physics_2006}%
  \BibitemOpen
  \bibfield  {author} {\bibinfo {author} {\bibfnamefont {S.~M.}\ \bibnamefont
  {Sze}}\ and\ \bibinfo {author} {\bibfnamefont {K.~K.}\ \bibnamefont {Ng}},\
  }\href@noop {} {\emph {\bibinfo {title} {Physics of {Semiconductor}
  {Devices}}}}\ (\bibinfo  {publisher} {John Wiley \& Sons},\ \bibinfo {year}
  {2006})\ \bibinfo {note} {google-Books-ID: o4unkmHBHb8C}\BibitemShut
  {NoStop}%
\bibitem [{\citenamefont {Nashima}\ \emph {et~al.}(2001)\citenamefont
  {Nashima}, \citenamefont {Morikawa}, \citenamefont {Takata},\ and\
  \citenamefont {Hangyo}}]{nashima_measurement_2001}%
  \BibitemOpen
  \bibfield  {author} {\bibinfo {author} {\bibfnamefont {S.}~\bibnamefont
  {Nashima}}, \bibinfo {author} {\bibfnamefont {O.}~\bibnamefont {Morikawa}},
  \bibinfo {author} {\bibfnamefont {K.}~\bibnamefont {Takata}}, \ and\ \bibinfo
  {author} {\bibfnamefont {M.}~\bibnamefont {Hangyo}},\ }\bibfield  {title}
  {\enquote {\bibinfo {title} {Measurement of optical properties of highly
  doped silicon by terahertz time domain reflection spectroscopy},}\ }\href
  {\doibase 10.1063/1.1413498} {\bibfield  {journal} {\bibinfo  {journal}
  {Applied Physics Letters}\ }\textbf {\bibinfo {volume} {79}},\ \bibinfo
  {pages} {3923--3925} (\bibinfo {year} {2001})}\BibitemShut {NoStop}%
\bibitem [{\citenamefont {Holm}\ and\ \citenamefont
  {Champlin}(1968)}]{holm_microwave_1968}%
  \BibitemOpen
  \bibfield  {author} {\bibinfo {author} {\bibfnamefont {J.~D.}\ \bibnamefont
  {Holm}}\ and\ \bibinfo {author} {\bibfnamefont {K.~S.}\ \bibnamefont
  {Champlin}},\ }\bibfield  {title} {\enquote {\bibinfo {title} {Microwave
  {Conductivity} of {Silicon} and {Germanium}},}\ }\href {\doibase
  10.1063/1.1655744} {\bibfield  {journal} {\bibinfo  {journal} {Journal of
  Applied Physics}\ }\textbf {\bibinfo {volume} {39}},\ \bibinfo {pages}
  {275--284} (\bibinfo {year} {1968})}\BibitemShut {NoStop}%
\bibitem [{\citenamefont {Michelotti}\ \emph {et~al.}(2009)\citenamefont
  {Michelotti}, \citenamefont {Dominici}, \citenamefont {Descrovi},
  \citenamefont {Danz},\ and\ \citenamefont
  {Menchini}}]{michelotti_thickness_2009}%
  \BibitemOpen
  \bibfield  {author} {\bibinfo {author} {\bibfnamefont {F.}~\bibnamefont
  {Michelotti}}, \bibinfo {author} {\bibfnamefont {L.}~\bibnamefont
  {Dominici}}, \bibinfo {author} {\bibfnamefont {E.}~\bibnamefont {Descrovi}},
  \bibinfo {author} {\bibfnamefont {N.}~\bibnamefont {Danz}}, \ and\ \bibinfo
  {author} {\bibfnamefont {F.}~\bibnamefont {Menchini}},\ }\bibfield  {title}
  {\enquote {\bibinfo {title} {Thickness dependence of surface plasmon
  polariton dispersion in transparent conducting oxide films at 1.55 $\mu$m},}\
  }\href {\doibase 10.1364/OL.34.000839} {\bibfield  {journal} {\bibinfo
  {journal} {Optics Letters}\ }\textbf {\bibinfo {volume} {34}},\ \bibinfo
  {pages} {839--841} (\bibinfo {year} {2009})}\BibitemShut {NoStop}%
\bibitem [{\citenamefont {Noginov}\ \emph {et~al.}(2011)\citenamefont
  {Noginov}, \citenamefont {Gu}, \citenamefont {Livenere}, \citenamefont {Zhu},
  \citenamefont {Pradhan}, \citenamefont {Mundle}, \citenamefont {Bahoura},
  \citenamefont {Barnakov},\ and\ \citenamefont
  {Podolskiy}}]{noginov_transparent_2011}%
  \BibitemOpen
  \bibfield  {author} {\bibinfo {author} {\bibfnamefont {M.~A.}\ \bibnamefont
  {Noginov}}, \bibinfo {author} {\bibfnamefont {L.}~\bibnamefont {Gu}},
  \bibinfo {author} {\bibfnamefont {J.}~\bibnamefont {Livenere}}, \bibinfo
  {author} {\bibfnamefont {G.}~\bibnamefont {Zhu}}, \bibinfo {author}
  {\bibfnamefont {A.~K.}\ \bibnamefont {Pradhan}}, \bibinfo {author}
  {\bibfnamefont {R.}~\bibnamefont {Mundle}}, \bibinfo {author} {\bibfnamefont
  {M.}~\bibnamefont {Bahoura}}, \bibinfo {author} {\bibfnamefont {Y.~A.}\
  \bibnamefont {Barnakov}}, \ and\ \bibinfo {author} {\bibfnamefont {V.~A.}\
  \bibnamefont {Podolskiy}},\ }\bibfield  {title} {\enquote {\bibinfo {title}
  {Transparent conductive oxides: {Plasmonic} materials for telecom
  wavelengths},}\ }\href {\doibase 10.1063/1.3604792} {\bibfield  {journal}
  {\bibinfo  {journal} {Applied Physics Letters}\ }\textbf {\bibinfo {volume}
  {99}},\ \bibinfo {pages} {021101} (\bibinfo {year} {2011})}\BibitemShut
  {NoStop}%
\bibitem [{\citenamefont {Naik}, \citenamefont {Shalaev},\ and\ \citenamefont
  {Boltasseva}(2013)}]{naik_alternative_2013}%
  \BibitemOpen
  \bibfield  {author} {\bibinfo {author} {\bibfnamefont {G.~V.}\ \bibnamefont
  {Naik}}, \bibinfo {author} {\bibfnamefont {V.~M.}\ \bibnamefont {Shalaev}}, \
  and\ \bibinfo {author} {\bibfnamefont {A.}~\bibnamefont {Boltasseva}},\
  }\bibfield  {title} {\enquote {\bibinfo {title} {Alternative {Plasmonic}
  {Materials}: {Beyond} {Gold} and {Silver}},}\ }\href {\doibase
  10.1002/adma.201205076} {\bibfield  {journal} {\bibinfo  {journal} {Advanced
  Materials}\ }\textbf {\bibinfo {volume} {25}},\ \bibinfo {pages} {3264--3294}
  (\bibinfo {year} {2013})}\BibitemShut {NoStop}%
\bibitem [{\citenamefont {Amin}\ \emph
  {et~al.}(2018{\natexlab{c}})\citenamefont {Amin}, \citenamefont {Tahersima},
  \citenamefont {Ma}, \citenamefont {Suer}, \citenamefont {Liu}, \citenamefont
  {Dalir},\ and\ \citenamefont {Sorger}}]{amin_low-loss_2018}%
  \BibitemOpen
  \bibfield  {author} {\bibinfo {author} {\bibfnamefont {R.}~\bibnamefont
  {Amin}}, \bibinfo {author} {\bibfnamefont {M.~H.}\ \bibnamefont {Tahersima}},
  \bibinfo {author} {\bibfnamefont {Z.}~\bibnamefont {Ma}}, \bibinfo {author}
  {\bibfnamefont {C.}~\bibnamefont {Suer}}, \bibinfo {author} {\bibfnamefont
  {K.}~\bibnamefont {Liu}}, \bibinfo {author} {\bibfnamefont {H.}~\bibnamefont
  {Dalir}}, \ and\ \bibinfo {author} {\bibfnamefont {V.~J.}\ \bibnamefont
  {Sorger}},\ }\bibfield  {title} {\enquote {\bibinfo {title} {Low-loss tunable
  1d {ITO}-slot photonic crystal nanobeam cavity},}\ }\href {\doibase
  10.1088/2040-8986/aab8bf} {\bibfield  {journal} {\bibinfo  {journal} {Journal
  of Optics}\ }\textbf {\bibinfo {volume} {20}},\ \bibinfo {pages} {054003}
  (\bibinfo {year} {2018}{\natexlab{c}})}\BibitemShut {NoStop}%
\bibitem [{\citenamefont {Amin}\ \emph {et~al.}(2017)\citenamefont {Amin},
  \citenamefont {Suer}, \citenamefont {Ma}, \citenamefont {Sarpkaya},
  \citenamefont {Khurgin}, \citenamefont {Agarwal},\ and\ \citenamefont
  {Sorger}}]{amin_deterministic_2017}%
  \BibitemOpen
  \bibfield  {author} {\bibinfo {author} {\bibfnamefont {R.}~\bibnamefont
  {Amin}}, \bibinfo {author} {\bibfnamefont {C.}~\bibnamefont {Suer}}, \bibinfo
  {author} {\bibfnamefont {Z.}~\bibnamefont {Ma}}, \bibinfo {author}
  {\bibfnamefont {I.}~\bibnamefont {Sarpkaya}}, \bibinfo {author}
  {\bibfnamefont {J.~B.}\ \bibnamefont {Khurgin}}, \bibinfo {author}
  {\bibfnamefont {R.}~\bibnamefont {Agarwal}}, \ and\ \bibinfo {author}
  {\bibfnamefont {V.~J.}\ \bibnamefont {Sorger}},\ }\bibfield  {title}
  {\enquote {\bibinfo {title} {A deterministic guide for material and mode
  dependence of on-chip electro-optic modulator performance},}\ }\href
  {\doibase 10.1016/j.sse.2017.06.024} {\bibfield  {journal} {\bibinfo
  {journal} {Solid-State Electronics}\ }\textbf {\bibinfo {volume} {136}},\
  \bibinfo {pages} {92--101} (\bibinfo {year} {2017})}\BibitemShut {NoStop}%
\bibitem [{\citenamefont {LeCun}\ and\ \citenamefont
  {Cortes}(2010)}]{lecun-mnisthandwrittendigit-2010}%
  \BibitemOpen
  \bibfield  {author} {\bibinfo {author} {\bibfnamefont {Y.}~\bibnamefont
  {LeCun}}\ and\ \bibinfo {author} {\bibfnamefont {C.}~\bibnamefont {Cortes}},\
  }\href {http://yann.lecun.com/exdb/mnist/} {\enquote {\bibinfo {title}
  {{MNIST} handwritten digit database},}\ }\bibinfo {howpublished}
  {http://yann.lecun.com/exdb/mnist/} (\bibinfo {year} {2010})\BibitemShut
  {NoStop}%
\bibitem [{\citenamefont {Chollet}\ \emph {et~al.}(2015)\citenamefont {Chollet}
  \emph {et~al.}}]{chollet2015keras}%
  \BibitemOpen
  \bibfield  {author} {\bibinfo {author} {\bibfnamefont {F.}~\bibnamefont
  {Chollet}} \emph {et~al.},\ }\href@noop {} {\enquote {\bibinfo {title}
  {Keras},}\ }\bibinfo {howpublished} {\url{https://keras.io}} (\bibinfo {year}
  {2015})\BibitemShut {NoStop}%
\bibitem [{\citenamefont {Abadi}\ \emph {et~al.}(2015)\citenamefont {Abadi},
  \citenamefont {Agarwal}, \citenamefont {Barham}, \citenamefont {Brevdo},
  \citenamefont {Chen}, \citenamefont {Citro}, \citenamefont {Corrado},
  \citenamefont {Davis}, \citenamefont {Dean}, \citenamefont {Devin},
  \citenamefont {Ghemawat}, \citenamefont {Goodfellow}, \citenamefont {Harp},
  \citenamefont {Irving}, \citenamefont {Isard}, \citenamefont {Jia},
  \citenamefont {Jozefowicz}, \citenamefont {Kaiser}, \citenamefont {Kudlur},
  \citenamefont {Levenberg}, \citenamefont {Man\'{e}}, \citenamefont {Monga},
  \citenamefont {Moore}, \citenamefont {Murray}, \citenamefont {Olah},
  \citenamefont {Schuster}, \citenamefont {Shlens}, \citenamefont {Steiner},
  \citenamefont {Sutskever}, \citenamefont {Talwar}, \citenamefont {Tucker},
  \citenamefont {Vanhoucke}, \citenamefont {Vasudevan}, \citenamefont
  {Vi\'{e}gas}, \citenamefont {Vinyals}, \citenamefont {Warden}, \citenamefont
  {Wattenberg}, \citenamefont {Wicke}, \citenamefont {Yu},\ and\ \citenamefont
  {Zheng}}]{tensorflow2015-whitepaper}%
  \BibitemOpen
  \bibfield  {author} {\bibinfo {author} {\bibfnamefont {M.}~\bibnamefont
  {Abadi}}, \bibinfo {author} {\bibfnamefont {A.}~\bibnamefont {Agarwal}},
  \bibinfo {author} {\bibfnamefont {P.}~\bibnamefont {Barham}}, \bibinfo
  {author} {\bibfnamefont {E.}~\bibnamefont {Brevdo}}, \bibinfo {author}
  {\bibfnamefont {Z.}~\bibnamefont {Chen}}, \bibinfo {author} {\bibfnamefont
  {C.}~\bibnamefont {Citro}}, \bibinfo {author} {\bibfnamefont {G.~S.}\
  \bibnamefont {Corrado}}, \bibinfo {author} {\bibfnamefont {A.}~\bibnamefont
  {Davis}}, \bibinfo {author} {\bibfnamefont {J.}~\bibnamefont {Dean}},
  \bibinfo {author} {\bibfnamefont {M.}~\bibnamefont {Devin}}, \bibinfo
  {author} {\bibfnamefont {S.}~\bibnamefont {Ghemawat}}, \bibinfo {author}
  {\bibfnamefont {I.}~\bibnamefont {Goodfellow}}, \bibinfo {author}
  {\bibfnamefont {A.}~\bibnamefont {Harp}}, \bibinfo {author} {\bibfnamefont
  {G.}~\bibnamefont {Irving}}, \bibinfo {author} {\bibfnamefont
  {M.}~\bibnamefont {Isard}}, \bibinfo {author} {\bibfnamefont
  {Y.}~\bibnamefont {Jia}}, \bibinfo {author} {\bibfnamefont {R.}~\bibnamefont
  {Jozefowicz}}, \bibinfo {author} {\bibfnamefont {L.}~\bibnamefont {Kaiser}},
  \bibinfo {author} {\bibfnamefont {M.}~\bibnamefont {Kudlur}}, \bibinfo
  {author} {\bibfnamefont {J.}~\bibnamefont {Levenberg}}, \bibinfo {author}
  {\bibfnamefont {D.}~\bibnamefont {Man\'{e}}}, \bibinfo {author}
  {\bibfnamefont {R.}~\bibnamefont {Monga}}, \bibinfo {author} {\bibfnamefont
  {S.}~\bibnamefont {Moore}}, \bibinfo {author} {\bibfnamefont
  {D.}~\bibnamefont {Murray}}, \bibinfo {author} {\bibfnamefont
  {C.}~\bibnamefont {Olah}}, \bibinfo {author} {\bibfnamefont {M.}~\bibnamefont
  {Schuster}}, \bibinfo {author} {\bibfnamefont {J.}~\bibnamefont {Shlens}},
  \bibinfo {author} {\bibfnamefont {B.}~\bibnamefont {Steiner}}, \bibinfo
  {author} {\bibfnamefont {I.}~\bibnamefont {Sutskever}}, \bibinfo {author}
  {\bibfnamefont {K.}~\bibnamefont {Talwar}}, \bibinfo {author} {\bibfnamefont
  {P.}~\bibnamefont {Tucker}}, \bibinfo {author} {\bibfnamefont
  {V.}~\bibnamefont {Vanhoucke}}, \bibinfo {author} {\bibfnamefont
  {V.}~\bibnamefont {Vasudevan}}, \bibinfo {author} {\bibfnamefont
  {F.}~\bibnamefont {Vi\'{e}gas}}, \bibinfo {author} {\bibfnamefont
  {O.}~\bibnamefont {Vinyals}}, \bibinfo {author} {\bibfnamefont
  {P.}~\bibnamefont {Warden}}, \bibinfo {author} {\bibfnamefont
  {M.}~\bibnamefont {Wattenberg}}, \bibinfo {author} {\bibfnamefont
  {M.}~\bibnamefont {Wicke}}, \bibinfo {author} {\bibfnamefont
  {Y.}~\bibnamefont {Yu}}, \ and\ \bibinfo {author} {\bibfnamefont
  {X.}~\bibnamefont {Zheng}},\ }\href {https://www.tensorflow.org/} {\enquote
  {\bibinfo {title} {{TensorFlow}: Large-scale machine learning on
  heterogeneous systems},}\ } (\bibinfo {year} {2015}),\ \bibinfo {note}
  {software available from tensorflow.org}\BibitemShut {NoStop}%
\bibitem [{\citenamefont {Duchi}, \citenamefont {Hazan},\ and\ \citenamefont
  {Singer}(2010)}]{Duchi:EECS-2010-24}%
  \BibitemOpen
  \bibfield  {author} {\bibinfo {author} {\bibfnamefont {J.}~\bibnamefont
  {Duchi}}, \bibinfo {author} {\bibfnamefont {E.}~\bibnamefont {Hazan}}, \ and\
  \bibinfo {author} {\bibfnamefont {Y.}~\bibnamefont {Singer}},\ }\href
  {http://www2.eecs.berkeley.edu/Pubs/TechRpts/2010/EECS-2010-24.html}
  {\enquote {\bibinfo {title} {Adaptive subgradient methods for online learning
  and stochastic optimization},}\ }\bibinfo {type} {Tech. Rep.}\ \bibinfo
  {number} {UCB/EECS-2010-24}\ (\bibinfo  {institution} {EECS Department,
  University of California, Berkeley},\ \bibinfo {year} {2010})\BibitemShut
  {NoStop}%
\bibitem [{\citenamefont {Wang}\ \emph {et~al.}(2018)\citenamefont {Wang},
  \citenamefont {Zhang}, \citenamefont {Stern}, \citenamefont {Lipson},\ and\
  \citenamefont {Loncar}}]{wang_nanophotonic_2018}%
  \BibitemOpen
  \bibfield  {author} {\bibinfo {author} {\bibfnamefont {C.}~\bibnamefont
  {Wang}}, \bibinfo {author} {\bibfnamefont {M.}~\bibnamefont {Zhang}},
  \bibinfo {author} {\bibfnamefont {B.}~\bibnamefont {Stern}}, \bibinfo
  {author} {\bibfnamefont {M.}~\bibnamefont {Lipson}}, \ and\ \bibinfo {author}
  {\bibfnamefont {M.}~\bibnamefont {Loncar}},\ }\bibfield  {title} {\enquote
  {\bibinfo {title} {Nanophotonic lithium niobate electro-optic modulators},}\
  }\href {\doibase 10.1364/OE.26.001547} {\bibfield  {journal} {\bibinfo
  {journal} {Optics Express}\ }\textbf {\bibinfo {volume} {26}},\ \bibinfo
  {pages} {1547--1555} (\bibinfo {year} {2018})}\BibitemShut {NoStop}%
\bibitem [{\citenamefont {Patel}\ \emph {et~al.}(2018)\citenamefont {Patel},
  \citenamefont {Parvizi}, \citenamefont {Ben-Hamida}, \citenamefont
  {Rolland},\ and\ \citenamefont {Plant}}]{patel_frequency_2018}%
  \BibitemOpen
  \bibfield  {author} {\bibinfo {author} {\bibfnamefont {D.}~\bibnamefont
  {Patel}}, \bibinfo {author} {\bibfnamefont {M.}~\bibnamefont {Parvizi}},
  \bibinfo {author} {\bibfnamefont {N.}~\bibnamefont {Ben-Hamida}}, \bibinfo
  {author} {\bibfnamefont {C.}~\bibnamefont {Rolland}}, \ and\ \bibinfo
  {author} {\bibfnamefont {D.~V.}\ \bibnamefont {Plant}},\ }\bibfield  {title}
  {\enquote {\bibinfo {title} {Frequency response of dual-drive silicon
  photonic modulators with coupling between electrodes},}\ }\href {\doibase
  10.1364/OE.26.008904} {\bibfield  {journal} {\bibinfo  {journal} {Optics
  Express}\ }\textbf {\bibinfo {volume} {26}},\ \bibinfo {pages} {8904}
  (\bibinfo {year} {2018})}\BibitemShut {NoStop}%
\bibitem [{\citenamefont {Streshinsky}\ \emph {et~al.}(2013)\citenamefont
  {Streshinsky}, \citenamefont {Ding}, \citenamefont {Liu}, \citenamefont
  {Novack}, \citenamefont {Yang}, \citenamefont {Ma}, \citenamefont {Tu},
  \citenamefont {Chee}, \citenamefont {Lim}, \citenamefont {Lo}, \citenamefont
  {Baehr-Jones},\ and\ \citenamefont {Hochberg}}]{streshinsky_low_2013}%
  \BibitemOpen
  \bibfield  {author} {\bibinfo {author} {\bibfnamefont {M.}~\bibnamefont
  {Streshinsky}}, \bibinfo {author} {\bibfnamefont {R.}~\bibnamefont {Ding}},
  \bibinfo {author} {\bibfnamefont {Y.}~\bibnamefont {Liu}}, \bibinfo {author}
  {\bibfnamefont {A.}~\bibnamefont {Novack}}, \bibinfo {author} {\bibfnamefont
  {Y.}~\bibnamefont {Yang}}, \bibinfo {author} {\bibfnamefont {Y.}~\bibnamefont
  {Ma}}, \bibinfo {author} {\bibfnamefont {X.}~\bibnamefont {Tu}}, \bibinfo
  {author} {\bibfnamefont {E.~K.~S.}\ \bibnamefont {Chee}}, \bibinfo {author}
  {\bibfnamefont {A.~E.-J.}\ \bibnamefont {Lim}}, \bibinfo {author}
  {\bibfnamefont {P.~G.-Q.}\ \bibnamefont {Lo}}, \bibinfo {author}
  {\bibfnamefont {T.}~\bibnamefont {Baehr-Jones}}, \ and\ \bibinfo {author}
  {\bibfnamefont {M.}~\bibnamefont {Hochberg}},\ }\bibfield  {title} {\enquote
  {\bibinfo {title} {Low power 50 {Gb}/s silicon traveling wave
  {Mach}-{Zehnder} modulator near 1300 nm},}\ }\href {\doibase
  10.1364/OE.21.030350} {\bibfield  {journal} {\bibinfo  {journal} {Optics
  Express}\ }\textbf {\bibinfo {volume} {21}},\ \bibinfo {pages} {30350}
  (\bibinfo {year} {2013})}\BibitemShut {NoStop}%
\bibitem [{\citenamefont {Li}\ \emph {et~al.}(2012)\citenamefont {Li},
  \citenamefont {Zheng}, \citenamefont {Thacker}, \citenamefont {Yao},
  \citenamefont {Luo}, \citenamefont {Shubin}, \citenamefont {Raj},
  \citenamefont {Cunningham},\ and\ \citenamefont
  {Krishnamoorthy}}]{li_40_2012}%
  \BibitemOpen
  \bibfield  {author} {\bibinfo {author} {\bibfnamefont {G.}~\bibnamefont
  {Li}}, \bibinfo {author} {\bibfnamefont {X.}~\bibnamefont {Zheng}}, \bibinfo
  {author} {\bibfnamefont {H.}~\bibnamefont {Thacker}}, \bibinfo {author}
  {\bibfnamefont {J.}~\bibnamefont {Yao}}, \bibinfo {author} {\bibfnamefont
  {Y.}~\bibnamefont {Luo}}, \bibinfo {author} {\bibfnamefont {I.}~\bibnamefont
  {Shubin}}, \bibinfo {author} {\bibfnamefont {K.}~\bibnamefont {Raj}},
  \bibinfo {author} {\bibfnamefont {J.~E.}\ \bibnamefont {Cunningham}}, \ and\
  \bibinfo {author} {\bibfnamefont {A.~V.}\ \bibnamefont {Krishnamoorthy}},\
  }\bibfield  {title} {\enquote {\bibinfo {title} {40 {Gb}/s thermally tunable
  {CMOS} ring modulator},}\ }in\ \href {\doibase 10.1109/GROUP4.2012.6324190}
  {\emph {\bibinfo {booktitle} {The 9th International Conference on Group IV
  Photonics GFP}}}\ (\bibinfo  {publisher} {IEEE},\ \bibinfo {address} {San
  Diego, CA, USA},\ \bibinfo {year} {2012})\ pp.\ \bibinfo {pages}
  {1--3}\BibitemShut {NoStop}%
\bibitem [{\citenamefont {Pantouvaki}\ \emph {et~al.}(2015)\citenamefont
  {Pantouvaki}, \citenamefont {Verheyen}, \citenamefont {De~Coster},
  \citenamefont {Lepage}, \citenamefont {Absil},\ and\ \citenamefont
  {Van~Campenhout}}]{pantouvaki_56gb/s_2015}%
  \BibitemOpen
  \bibfield  {author} {\bibinfo {author} {\bibfnamefont {M.}~\bibnamefont
  {Pantouvaki}}, \bibinfo {author} {\bibfnamefont {P.}~\bibnamefont
  {Verheyen}}, \bibinfo {author} {\bibfnamefont {J.}~\bibnamefont {De~Coster}},
  \bibinfo {author} {\bibfnamefont {G.}~\bibnamefont {Lepage}}, \bibinfo
  {author} {\bibfnamefont {P.}~\bibnamefont {Absil}}, \ and\ \bibinfo {author}
  {\bibfnamefont {J.}~\bibnamefont {Van~Campenhout}},\ }\bibfield  {title}
  {\enquote {\bibinfo {title} {56gb/s ring modulator on a 300mm silicon
  photonics platform},}\ }in\ \href {\doibase 10.1109/ECOC.2015.7341888} {\emph
  {\bibinfo {booktitle} {2015 {European} {Conference} on {Optical}
  {Communication} ({ECOC})}}}\ (\bibinfo  {publisher} {IEEE},\ \bibinfo
  {address} {Valencia, Spain},\ \bibinfo {year} {2015})\ pp.\ \bibinfo {pages}
  {1--3}\BibitemShut {NoStop}%
\bibitem [{\citenamefont {Timurdogan}\ \emph {et~al.}(2014)\citenamefont
  {Timurdogan}, \citenamefont {Sorace-Agaskar}, \citenamefont {Sun},
  \citenamefont {Shah~Hosseini}, \citenamefont {Biberman},\ and\ \citenamefont
  {Watts}}]{timurdogan_ultralow_2014}%
  \BibitemOpen
  \bibfield  {author} {\bibinfo {author} {\bibfnamefont {E.}~\bibnamefont
  {Timurdogan}}, \bibinfo {author} {\bibfnamefont {C.~M.}\ \bibnamefont
  {Sorace-Agaskar}}, \bibinfo {author} {\bibfnamefont {J.}~\bibnamefont {Sun}},
  \bibinfo {author} {\bibfnamefont {E.}~\bibnamefont {Shah~Hosseini}}, \bibinfo
  {author} {\bibfnamefont {A.}~\bibnamefont {Biberman}}, \ and\ \bibinfo
  {author} {\bibfnamefont {M.~R.}\ \bibnamefont {Watts}},\ }\bibfield  {title}
  {\enquote {\bibinfo {title} {An ultralow power athermal silicon modulator},}\
  }\href {\doibase 10.1038/ncomms5008} {\bibfield  {journal} {\bibinfo
  {journal} {Nature Communications}\ }\textbf {\bibinfo {volume} {5}},\
  \bibinfo {pages} {4008} (\bibinfo {year} {2014})}\BibitemShut {NoStop}%
\bibitem [{\citenamefont {Kim}\ and\ \citenamefont
  {Kim}(2016)}]{kim_silicon_2016}%
  \BibitemOpen
  \bibfield  {author} {\bibinfo {author} {\bibfnamefont {J.-S.}\ \bibnamefont
  {Kim}}\ and\ \bibinfo {author} {\bibfnamefont {J.~T.}\ \bibnamefont {Kim}},\
  }\bibfield  {title} {\enquote {\bibinfo {title} {Silicon electro-optic
  modulator based on an {ITO}-integrated tunable directional coupler},}\ }\href
  {\doibase 10.1088/0022-3727/49/7/075101} {\bibfield  {journal} {\bibinfo
  {journal} {Journal of Physics D: Applied Physics}\ }\textbf {\bibinfo
  {volume} {49}},\ \bibinfo {pages} {075101} (\bibinfo {year}
  {2016})}\BibitemShut {NoStop}%
\end{thebibliography}
\end{document}